\RequirePackage{fix-cm}
\documentclass[twocolumn,epjc3]{svjour3}
\smartqed  
\RequirePackage{graphicx}
\usepackage{epsfig}
\usepackage{amssymb}
\usepackage{bm} 
\usepackage[usenames]{color}
\usepackage{bbding}
\usepackage{slashed}
\usepackage{array}
\usepackage{cite}  
\newcommand{\Tr}{\mbox{Tr}}
\newcommand{\vphi}{\varphi}
\newcommand{\Lagr}{{\cal L}}

\newcommand{\omegaPP}{\bar{\omega}^{\perp}}
\newcommand{\omegaPL}{\bar{\omega}^{\parallel}}

\newcommand{\pard}{\partial}
\newcommand{\BB}{\mbox{\boldmath$B$}}
\newcommand{\BW}{\mbox{\boldmath$W$}}
\newcommand{\BV}{\mbox{\boldmath$V$}}

\newcommand{\BWslash}{\!\not{\!\! \BW}}

\newcommand{\BBslash}{\!\not{\!\! \BB}}

\newcommand{\BVslash}{\!\not{\! \BV}}
\newcommand{\cL}{{\cal L}}

\newcommand{\Comm}[2]{\mbox{$[$} #1, #2\mbox{$]$}}
\newcommand{\diag}{\mbox{diag}}

\newcommand{\unit}[1]{\;\mathrm{#1}}
\journalname{Eur. Phys. J. C}
\begin{document}

\title{
A case study about the mass exclusion limits for the BSM vector resonances
with the direct couplings to the third quark generation
}


\author{Mikul\'{a}\v{s} Gintner\thanksref{e1,addr1,addr3}
        \and
        Josef Jur\'{a}\v{n}\thanksref{e2,addr1,addr2} 
}

\thankstext{e1}{e-mail: gintner@fyzika.uniza.sk}
\thankstext{e2}{e-mail: josef.juran@utef.cvut.cz}


\institute{Institute of Experimental and Applied Physics,
Czech Technical University in Prague, Husova 240/5, 110 00 Prague, 
Czech Republic \label{addr1}
           \and
           Institute of Physics, Silesian University in Opava,
Bezru\v{c}ovo n\'{a}m. 13, 746 01 Opava, Czech Republic \label{addr2}
           \and
           Physics Department, University of \v{Z}ilina,
Univerzitn\'{a} 1, 010 26 \v{Z}ilina, Slovakia
\label{addr3} }

\date{Received: date / Accepted: date}

\maketitle

\begin{abstract}
The upper bounds that the LHC measurements searching for heavy resonances 
beyond the Standard model set on the resonance production cross sections
are not universal. 
They depend on various characteristics of the resonance under consideration,
like its mass, spin, and its interaction pattern. Their validity is
also limited by the assumptions and approximations applied to
their calculations. The bounds are typically used to derive the mass exclusion
limits for the new resonances. In our work, we address some of the issues
that emerge when deriving the mass exclusion limits for the strongly coupled composite
$SU(2)_{L+R}$ vector resonance triplet which would interact directly to the third
quark generation only. We investigate the restrictions on the applicability
of the generally used limit-obtaining procedure to this particular type
of vector resonances. We demonstrate that, in this case, it is necessary to consider 
the bottom quark partonic contents of the proton.
Eventually, we find the mass exclusion limits for this resonance triplet
for some representative subsets of the parameter space.

\end{abstract}

\section{Introduction}
\label{sec:intro}
The existence of new particles, complementing the established Standard model (SM) spectrum, 
is predicted by all major scenarios of the SM extension. Should 
they be the supersymmetric partners to the SM fields or the composite resonances 
of new strong interactions, the discovery of the new particle(s) would provide 
undeniable evidence for beyond the SM physics. No wonder that the search for 
them has its rightful place in the ATLAS and CMS Collaboration's activities. 
Nevertheless,  despite all their effort, no new particle has been discovered so 
far. Actually, not even a significant disagreement with the SM predictions has 
been observed yet.

Facing the absence of the positive experimental input the available data can be 
used to set the exclusion limits on the parameters of the candidate BSM 
theories. First of all, the data can be translated into the excluded values of 
the masses of the sought-after new particles. Indeed, both Collaborations are 
making an effort to process their measurements into the form that can be used to 
establish the mass exclusion limits for the new particles of various kinds. 

Obtaining the bound and, subsequently, the mass exclusion limit for the 
resonance of a particular BSM theory is a challenging task that requires the 
contributions of the experimental as well as theoretical communities.
Since there is plenty of candidate theories with new particles of different 
properties, both communities seek to make the procedure as simple 
and general as possible.
However, there are principal restrictions on how model-independent the analysis 
can become. They result from the specific properties of the sought-after 
resonance, like its spin or the absence of a certain decay channel. To calculate 
the mass exclusion limits correctly all significant
model-imposed assumptions have to be identified and taken into account properly.
Any simplifications, introduced in the calculations of the model cross sections, 
are welcomed. However, one has to remember that the simplifying assumptions can 
result in the reduction of the applicability of the obtained exclusion limits. 

In their bounds producing analyses, the ATLAS and CMS Collaborations have 
focused on the on-shell direct production of the new particles in the LHC 
proton-proton collisions. There, the resonances are searched
for in their various two-particle decay channels. The absence of the 
statistically significant deviation from the SM prediction in a given decay 
channel is then being translated into the upper bound on the production cross section 
of the BSM resonance multiplied by the relevant branching ratio. 

In order to work out the mass exclusion limit for a particular BSM model 
resonance theorists have to deliver the model's cross sections that are
to be compared with the experimental upper bounds for the individual channels. 
If such a cross section exceeds the bound the resonance (and, thus, the model) 
is considered as being excluded by the experiment. 
When the resonance mass is a free parameter of the model, the regions of the 
excluded mass values can be established in this way.

In the strongly-interacting extensions of the SM, multiple new resonances
emerge as the bound states of new strongly-interacting fundamental fields.
Since the forces responsible for the creation of the resonances are non-perturbative, 
a very limited information about the resonances can be obtained from the first 
principles. Thus, it is highly desirable to develop and use the effective description
of the BSM bound state phenomenology; particularly, it would be the phenomenology 
of the bound states that might be observable at the LHC.
Naturally, these usually include the lightest bound states of the given theory.
While, in some scenarios, the spin-0 and spin-1 resonances are expected to be the lightest 
bound states~\cite{ref1,ref2,ref3,ref4}, there are also light fermionic resonances
predicted in others, see, e.g.,~\cite{HiggsInMultiplet,ContinoResonancesInCmpHiggs}.

Typical representatives of the strongly-interacting theories are the
Technicolor model~\cite{TC} and its extensions~\cite{ETC,TCWalking,TCTopcolor}.
More recent extra-dimensional theories~\cite{ExtraDim} predict the Kaluza-Klein 
towers of new resonances of which the lowest lying ones might be
observed at the LHC. The attractiveness of this development lies in
the dual-description relation between the extra-dimensional
weakly-interacting theories and the strongly-interacting models in
four dimensions~\cite{Maldacena}. Closely related are the scenarios in which
the Higgs appears as a composite pseudo-Goldstone boson of a strongly-coupled theory.
Simple four-dimensional models, like the three-site~\cite{ref5}, the four-site~\cite{ref6} 
and the effective composite Higgs model~\cite{ref7} can be used to characterize 
the main features of the emerging phenomenology.

In this paper, we focus on the strongly-interacting 
BSM vector resonances with a particular pattern of the interactions with 
the SM fermions: they couple directly to the top and bottom quarks only.
Our ambition is to derive the exclusion limits for this kind of scenario
and to understand the impact of its characteristic features on the applicability of 
the used simplifications.
For this purpose, we use the effective Lagrangian that we developed and studied
in~\cite{tBESSprd11,tBESSepjc13,tBESSepjc16,tBESSapp18} (the tBESS model).

Intuitively, it is natural to suspect that, due to its extraordinary mass,
the top quark would play an outstanding
role in the context of the new strong physics. The top mass
being surprisingly close to the scale of the electroweak symmetry breaking (ESB)
might be generated by the same mechanism as the masses of the electroweak gauge 
bosons~\cite{snowmass,Han4}.
This happens, for example, for the vector
$\rho_T$ resonance of the Extended Technicolor~\cite{ETC}.
Yet, the mechanism behind the top mass could differ from that of ESB
as it happens in the Topcolor Assisted Technicolor~\cite{TCTopcolor}.
All these possibilities can result in the interactions of the vector
resonances with the top quark that are different from the interactions with 
the light SM fermions.
The neutral component of the new vector triplet
can also mimic couplings of a $Z'$ spin-1 resonance~\cite{Han3,He} 
which has large couplings to $t$ and $b$, vanishing
couplings to $W$, $Z$ and very small couplings to fermions of the
first two generations. Regarding the enlarged
couplings of the third generation quarks to the vector resonance triplet, 
similar features can be found in the Composite Higgs models~\cite{
CompositeScalar-GGduality,CHS,Xie2019}
employing the idea of partial compositeness~\cite{Kaplan}.
Since these models predict also new fermion resonances our effective description
would apply to the cases when the effects of the fermions can be neglected.
It is true that the Higgs boson in these models is a Goldstone boson rather than
a generic singlet. Nevertheless, this does not have to change its low-energy dynamics 
dramatically.

Note that the effective description of the possible LHC phenomenology provided
by the tBESS model is not necessarily in the rigorous one-to-one correspondence with 
any of the theories mentioned above. It can rather provide the effective description 
of the LHC phenomenology for proper subregions of the parameter space of various candidate
theories. Thus, for example, while the richness of the resonance spectrum of many 
strongly-interacting theories far exceeds the particle contents of the tBESS model,
this model might provide a satisfactory description of the LHC phenomenology
for the parameter values when the majority of the resonances can be integrated out.
In this way we can build a simplifying bridge for the application of the 
LHC experimental findings to the existing theories. The examples of 
this approach can be found in the literature, some of them even sharing certain features 
of the tBESS model.
The effective description of the situation when the third quark generation couples 
extraordinarily to the new scalar and vector strong resonances was
studied also in~\cite{Han1,Han2,Han3}. Another example of this approach when
two, vector and axial-vector, $SU(2)$ triplets of the strong resonances were considered
can be found in~\cite{DeCurtis}. A ``composite'' scalar-vector system at the LHC
was also studied in~\cite{Hernandez}. The phenomenology of a broad vector resonance
in a more general formalism was studied in~\cite{Chivukula}. 
The broad vector resonance case decaying mainly to the third-generation quarks 
was also studied recently in~\cite{Xie2019}.

In the tBESS effective model the Higgs sector 
is based on the non-linear sigma model with the 125-GeV 
$SU(2)_{L+R}$ scalar singlet complementing its non-linear triplet of the 
Nambu-Goldstone bosons. The vector resonances are present in the Lagrangian as 
an $SU(2)_{L+R}$ triplet. This setup fits the situation when the global 
$SU(2)_L\times SU(2)_R$ symmetry is broken down to $SU(2)_{L+R}$.

The vector triplet is introduced as a gauge field via the hidden local symmetry 
approach~\cite{HLS}. Thus, the mass eigenstate representation of the vector 
resonance contains the admixture of EW gauge bosons.
The gauge sector of this effective description is equivalent to the gauge sector 
of highly-deconstructed Higgsless model with three sites~\cite{ref5}. 

There are two specific features of the model that should be called attention to. 
First, the 
mass and decay widths of the vector resonances are entangled with the model's 
couplings. The vector resonance total width grows quite quickly with the 
resonance mass. The masses of the neutral and charged vector resonances are 
virtually degenerate when the resonance coupling $g''$ is much bigger than the 
$SU(2)_L\times U(1)_Y$ gauge couplings ($g$, $g'$).

Secondly, as advertised above, the direct interactions of the vector triplet
with the third generation quarks only are admitted in the fermion sector.
These interactions grow with $g''$ and can be introduced as flavor and chirality 
dependent, with no other interactions of the vector triplet with fermions 
in the flavor basis\footnote{Other sectors include self-interactions 
and the interactions with the SM EW gauge bosons and the Higgs boson.}. However, 
the couplings of the vector resonance to the light SM fermions emerge 
in the mass eigenstate basis
due to the mixing of the gauge bosons. These interactions (referred to as 
\textit{indirect} couplings) are universal and suppressed by $1/g''$. 
Thanks to the mixing-induced 
couplings, it is possible to produce the vector resonances also in the 
light-quark Drell-Yan processes at the LHC.

The experimental upper bounds, provided by the ATLAS and CMS Collaborations, are 
based on the narrow width resonance assumption.
This assumption has its obvious benefits when deriving the experimental limits 
as well as for the calculation of the corresponding theoretical predictions. On 
the other hand, it restricts the scope of the method. 
It can be seen in our model. There are mass regions in the parametric space 
where the resonances are not so narrow and where the ratio 
exceeds the rule-of-thumb value of $10\%$. We identify the 
regions of the parameter space where the method can be applied to the vector 
triplet under consideration.

In our paper~\cite{tBESSapp18}, we did the analysis of the exclusion limits for 
our effective model with no direct couplings of the vector resonance triplet to 
the SM fermions. The universality of the indirect couplings justified the 
up-down quark only approximation of the proton partonic content under which the 
predictions of our model were calculated.

In the present work, the question about the role of the 
b-quark partons in the vector boson production reappears due to the direct
interactions between the vector triplet and the bottom quarks. 
Once more, one is tempted to ignore the 
tiny presence of the b-quarks in the proton, as we did in~\cite{tBESSapp18}. 
However, it will be shown that the direct couplings of the vector resonance with 
the bottom quark can even overwhelm the universal indirect 
interactions with fermions and, thus, compensate for the deficiency of the 
bottom quarks in the proton. In this paper, the first quark generation approximation 
used in~\cite{tBESSapp18} is upgraded to the all-sea-quark calculation.

In this paper, we establish the mass exclusion limits on the neutral and charged 
vector resonances of our model and observe their behavior. 
Because of the large dimensionality of the parameter space the full exclusion limit
analysis would be a difficult task. Therefore, 
we determine the exclusion limits for selected subsets of the parameter space only.
The limits are based 
on the most recent upper bounds on the LHC production cross sections times 
branching ratios published by the ATLAS and CMS Collaborations. To derive the 
exclusion limits we inspect and use all production mechanisms and decay channels 
relevant to our model for which the Collaborations published the upper 
bounds.

This paper is organized as follows. In Section~\ref{sec:effLagrangian}, our 
model is briefly introduced and its phenomenology concerning its decay widths 
and branching ratios studied. Section~\ref{sec:prodXS} is concerned with the 
calculations of the model cross sections that are to be compared with the 
experimental upper bounds. Section~\ref{sec:masslimits} contains the analysis 
and calculations of the exclusion mass limits for both cases of the model: 
without and with the direct interactions to the third quark generation. The 
results of our work are summarized in the Conclusions section.
\ref{app:Lagrangian} outlines the effective Lagrangian of our model.


\section{The effective Lagrangian and its phenomenology}
\label{sec:effLagrangian}
\subsection{The boson sectors}

The effective Lagrangian, we use in this paper, was studied in detail in our 
previous papers~\cite{tBESSprd11,tBESSepjc13,tBESSepjc16}. It can serve as an 
effective description of the LHC phenomenology of a hypothetical strongly 
interacting extension of the SM where the principal manifestation of this 
scenario would be the existence of a vector resonance triplet as a bound state 
of new strong interactions. The Lagrangian is built to respect the global 
$SU(2)_L\times SU(2)_R\times U(1)_{B-L}\times SU(2)_{HLS}$ symmetry of which the 
$SU(2)_L\times U(1)_Y\times SU(2)_{HLS}$ subgroup is also a local symmetry. The 
$SU(2)_{HLS}$ symmetry is an auxiliary gauge symmetry invoked to accommodate the 
$SU(2)_{L+R}$ triplet of new vector resonances. Each of the mentioned gauge 
groups is accompanied by its gauge coupling: $g$, $g'$, and $g''$ stand for 
$SU(2)_L$, $U(1)_Y$ and $SU(2)_{L+R}$, respectively. Beside the scalar singlet 
representing the 125 GeV Higgs boson and the hypothetical vector triplet, the 
effective Lagrangian is built out of the SM fields only. The effective 
Lagrangian itself can be found in~\cite{tBESSepjc16,tBESSepjc13,tBESSprd11}
and its basic structure is summarized in~\ref{app:Lagrangian}.

The way the vector resonance triplet is introduced into the effective Lagrangian 
implies the mixing between the resonance and electroweak gauge boson fields. To 
decipher the physical content of the Lagrangian the gauge fields have to be 
transformed from the flavor to mass eigenstate basis. Note that the Greek letter 
$\rho$ will denote the vector boson resonance fields in the mass eigenstate 
basis. Consequently, $\rho^\pm$ and $\rho^0$ stand for the charged and neutral 
members of the triplet, respectively.

When $g''\gg g,g'$ the masses of the charged and neutral vector resonances 
are virtually degenerate. The leading order formula for the 
vector resonance mass reads
\begin{equation}\label{eq:MassRho}
   M_\rho = \sqrt{\alpha}g''v/2,
\end{equation}
where $\alpha$ is a dimensionless free parameter emerging in the effective Lagrangian
and $v$ is the electroweak symmetry breaking scale.
Usually, $\alpha$ is traded off for $M_\rho$ so that the latter can serve as 
one of the free parameters of the model.
Our previous studies of the low-energy limits~\cite{tBESSprd11,tBESSepjc13} as well as
the Higgs-related limits and the
unitarity limits~\cite{tBESSepjc16} suggest that we should consider 
$M_\rho\geq 1\;\mbox{TeV}$ and $12\leq g''\leq 25$. 
Following the conventions used in the formulation
of our Lagrangian the naive perturbativity bound on $g''$
reads $g''/2\leq 4\pi$.

In this paper, we calculate processes with the direct production
of the vector resonances followed by their two-particle decays. 
In the boson sector, the contributing triple couplings of the model include 
$\rho WW$ and $\rho WZ$ triple interactions.
Their strengths are proportional to $1/g''$.
Note that the model contains neither of the $\rho ZZ$, $\rho Z\gamma$, $\rho\gamma\gamma$, and
$\rho W\gamma$ vertices.

Additional bosonic triple vertices that could play a role in setting the mass 
exclusion limits include the $\rho WH$ and $\rho ZH$ vertices. Both couplings 
are proportional to the $a_V-a_\rho$ difference and they are also suppressed by 
the factor $1/g''$. To a high precision, $a_V$ can be considered as a free 
pre-factor of the $HWW$ and $HZZ$ vertices. The $H\rho^0\rho^0$ and 
$H\rho^+\rho^-$ couplings are virtually proportional to $a_\rho$. The 
corresponding interaction Lagrangian, along with the calculations of the LHC 
experimental limits for $a_V$ and $a_\rho$, can be found in~\cite{tBESSepjc16}.

Throughout this paper, we set $a_V=1$ (the SM case) and 
$a_\rho=0$ (no Higgs-to-vector resonance coupling). These values are 
consistent with the experimentally preferred points of the parameter 
space~\cite{tBESSepjc16}. This choice that zeros the $H\rho^0\rho^0$ and $H\rho^+\rho^-$ 
vertices and sets the $HWW$ and $HZZ$ vertices to their SM form has no impact on
our analysis. Neither it affects 
the results through the $\rho WH$ and $\rho ZH$ vertices.

To the leading order in $g''$ the vector resonance partial decay widths to 
$W^+W^-$ and $W^\pm Z$ read
\begin{eqnarray}
   \Gamma_{\rho\rightarrow V_1V_2} &=& \Gamma_{\rho}\cdot
   \left[1+7(x_1^2+x_2^2)\right.
   \nonumber\\
   && \phantom{\Gamma_{\rho}[}\left.-26(x_1^4+x_2^4)-50x_1^2x_2^2+{\cal O}(x^6)\right],
\end{eqnarray}
where
\begin{equation}\label{eq:GammaWWandWZ}
   \Gamma_{\rho}= 
   \frac{M_\rho}{48\pi g^{\prime\prime 2}}\left(\frac{M_\rho}{v}\right)^4,
\end{equation}
$x_{1,2}=M_{1,2}/M_\rho$, and $V_i$ stands for either $W$ or $Z$.

The partial decay widths to the $WH$ and $WZ$ channels read
\begin{equation}\label{eq:GammaHW}
   \Gamma_{\rho^\pm\rightarrow W^\pm H} = 
   4\Gamma_\rho\cdot (a_\rho-a_V)^2\left[x_W^4+{\cal O}(x^6)\right],
\end{equation}
and
\begin{equation}\label{eq:GammaHZ}
   \Gamma_{\rho^0\rightarrow Z H} = 
   4\Gamma_\rho\cdot (a_\rho-a_V)^2\left[\left(2x_W^2-x_Z^2\right)^2+{\cal O}(x^6)\right],
\end{equation}
respectively. Within the 1 -- 3 TeV mass interval, the $WH$ and $WZ$ decay widths are four
to six orders of magnitude smaller than the $W^+W^-$ and $W^\pm Z$ ones. 

If there are no direct interactions of the vector resonance with fermions in the model,
the widths of the fermion-related decay channels of the vector resonance are also quite
negligible when compared to the $W^+W^-$ and $W^\pm Z$ decay channels. 
Thus, the total decay widths of the vector resonances, both neutral and
charged, can be well approximated by the expression~(\ref{eq:GammaWWandWZ}).

\begin{figure}[htb]
\centerline{
\includegraphics[width=8.5cm]{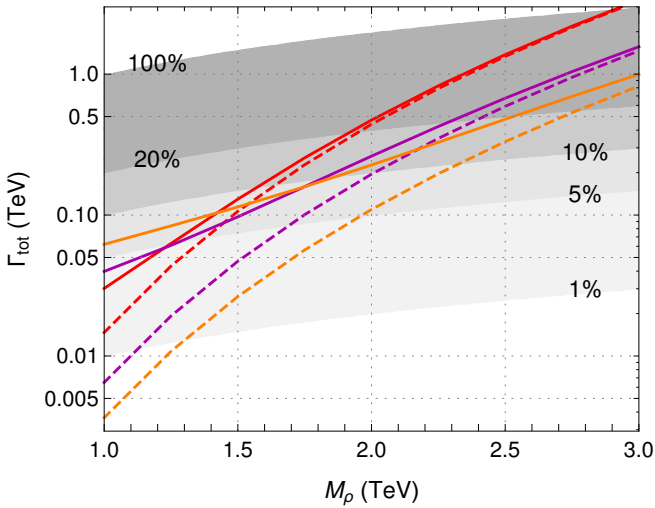}}
\caption{The total decay width of the vector resonance as a function of the 
resonance mass. Top to bottom, the dashed lines correspond to $g''=12$, $18$, 
and $24$, respectively, when there is no direct interaction of the vector 
resonance to fermions. The solid lines representing the case of the direct 
interactions with $b_{L,R}=-0.1$, and $p=1$ unite with their dashed 
$g''$-counterparts at the upper right corner of the graph. Different shadings of 
the background indicate areas with different value ranges of 
$\Gamma_\mathrm{tot}/M_\rho$. The curves do not distinguish between neutral and 
charged resonances.}
\label{Fig:TotalDecayWidth}
\end{figure}

In Fig.~\ref{Fig:TotalDecayWidth}, we depict how the vector resonance total width
depends on the resonance mass and $g''$ using the full tree-level formulas for the calculations. 
The dashed lines correspond to 
the no-direct-interaction total widths when $g''=12$, $18$, and $24$, respectively.
There is no visible distinction between the neutral and charged resonance curves
in this graph. Different shadings of the background indicate regions with
different values of the fatness of the resonance, $\Gamma_\mathrm{tot}/M_\rho$.
The graph also shows how the total width responds to the direct interactions.
However, this part will be discussed in the following subsection.

\subsection{The fermion sector}

The only fermions considered in our effective model are those of the SM. 
Thus, the model
describes the BSM situation when either non-SM fermionic fields are
much heavier than ${\cal O}(1)\unit{TeV}$ vector resonances or their
interactions make their existence irrelevant to our analysis.

Even if no direct couplings of the vector resonance
fields to fermions are introduced in the flavor eigenstate basis
the mixing between the vector resonance triplet and the electroweak gauge bosons
induces the couplings between the vector resonance and fermions. These ``indirect''
couplings are proportional to $1/g''$.

Nevertheless, the considered symmetry also admits the introduction of the direct 
interactions of the vector resonance with fermions. In fact, having grouped the 
right fermion fields into $SU(2)_R$ doublets the considered global symmetry of 
the model also allows for assigning different direct couplings to each of the 
$SU(2)_L$ and $SU(2)_R$ fermion doublets. 

We assume in our model that the flavor basis vector triplet couples directly to 
the quarks of the third generation only and to none of the other SM fermions. 
This assumption is motivated by the anticipated extraordinary role of the top 
quark (and, perhaps, bottom quark as well) in new physics related to strong 
electroweak symmetry breaking.  Similar interaction patterns can be found in 
various strong extensions of the SM including the partial compositeness and 
extra-dimensional scenarios. In our model, the free pre-factors for the 
couplings to the left and right top-bottom quark doublets are referred to as 
$b_L$ and $b_R$, respectively~\cite{tBESSprd11}.

In the mass basis, the direct interactions also contaminate the couplings of $W$ 
and $Z$ to the top and bottom quarks. The contamination is proportional to 
$b_{L,R}$. It results in the experimental restrictions on $b_{L,R}$ from the 
EW precision measurements~\cite{tBESSepjc13}.

In the most general case of our model, an additional free pre-factor $p$ is introduced
which serves to further modify the direct coupling to the right bottom quark. 
Assuming $0\leq p\leq 1$, the $p$
parameter can be used to suppress the direct coupling to the right bottom quark
relative to the direct coupling to the right top quark; $p=1$ leaves both interactions
equal, $p=0$ turns off the right bottom quark interaction completely and maximally breaks
the $SU(2)_R$ part of the Lagrangian symmetry down to $U(1)_{R3}$. Recall that
the $SU(2)_R$ symmetry is broken by the weak hypercharge interactions anyway and,
thus, the $SU(2)_R$ fermion doublets are not well justified once the global symmetry
gets gauged. 

This effective model was introduced and investigated 
in~\cite{tBESSepjc16,tBESSprd11,tBESSepjc13} as a modification of the so-called 
BESS model \cite{BESS}. We refer to our model as the ``top-BESS model'' (tBESS) in order to 
recall its relation to its predecessor as well as to stress the special role of 
the top quark (or the third quark generation) in it. As was shown 
in~\cite{tBESSprd11, tBESSepjc13}, these modifications also help relax some 
experimental restrictions burdening the original BESS model. Should we briefly 
summarize the conclusions reached in these studies, the absolute values of 
$b_{L,R}$ should not exceed $0.1$, roughly speaking. As far as $p$ is concerned, 
its most preferred value lies in the $0.2-0.3$ interval. However, the 
statistical preference of the interval with respect to any other value of $p$ 
between zero and one is weak.

Assuming massless fermions in the final state, the partial decay widths of 
the neutral vector resonance to the leptonic $\nu\nu$ and $\ell\ell$
channels read
\begin{equation}\label{eq:GammaNuNu}
   \Gamma_{\rho^0\rightarrow \nu\nu} = 8\Gamma_\rho\cdot
   \left(2x_W^2-x_Z^2\right)^2,
\end{equation}
and
\begin{equation}\label{eq:GammaLL}
   \Gamma_{\rho^0\rightarrow \ell\ell} = 8\Gamma_\rho\cdot
   \left[x_Z^4+4\left(x_Z^2-x_W^2\right)^2\right],
\end{equation}
respectively. Its partial decay widths to the light quark channels are
\begin{equation}\label{eq:GammaUU}
   \Gamma_{\rho^0\rightarrow uu} = 24\Gamma_\rho\cdot
   \frac{17x_Z^4+20x_W^4-28x_Z^2x_W^2}{9},
\end{equation}
and
\begin{equation}\label{eq:GammaDD}
   \Gamma_{\rho^0\rightarrow dd} = 24\Gamma_\rho\cdot
   \frac{5x_Z^4+20x_W^4-16x_Z^2x_W^2}{9},
\end{equation}
respectively. The corresponding partial decay widths of the charged vector resonance read
\begin{equation}\label{eq:GammaNuLUD}
   3\Gamma_{\rho^\pm\rightarrow \nu\ell} = \Gamma_{\rho^\pm\rightarrow ud} =
   48\Gamma_\rho\cdot x_W^4.
\end{equation}

The partial widths of the decay channels with the top and bottom quarks in the final
state are sensitive to the free parameters $b_{L,R}$ and $p$. 
They originate from the intertwining of the direct and indirect interactions.
The $tt$ partial width reads
\begin{eqnarray}\label{eq:GammaTT}
   \Gamma_{\rho^0\rightarrow tt} &=& \frac{M_{\rho}}{8\pi}\beta_t
   \left\{\left[(g_{\rho tt}^L)^2+(g_{\rho tt}^R)^2\right](1-x_t^2)\right.
   \nonumber\\
   &&\phantom{\frac{M_{\rho}}{8\pi}\beta_t}+\left.6g_{\rho tt}^L g_{\rho tt}^Rx_t^2\right\},
\end{eqnarray}
where $\beta_t=\sqrt{1-4x_t^2}$, $x_t=M_t/M_\rho$, and
\begin{eqnarray}
   g_{\rho tt}^L &=& -\frac{b_L}{4}g'' 
   + \frac{2}{g''}\left[\left(\frac{2}{3}-b_L \right)\frac{M_W^2}{v^2}
   +\frac{1}{3}\;\frac{M_Z^2}{v^2}\right],
   \label{eq:gttL}\\
   g_{\rho tt}^R &=& -\frac{b_R}{4}g'' 
   + \frac{2}{g''}\left(\frac{4}{3}-b_R \right)
   \left(\frac{M_Z^2}{v^2}-\frac{M_W^2}{v^2}\right).
   \label{eq:gttR}
\end{eqnarray}
To the leading order in $g''$, the partial width (\ref{eq:GammaTT}) reads
\begin{equation}\label{eq:GammaTTLO}
   \Gamma_{\rho^0\rightarrow tt} = \frac{3}{8}\Gamma_\rho\cdot
   g^{\prime\prime 4} \left[(b_L^2+b_R^2)\;
   x_v^4+{\cal O}(x^6)\right],
\end{equation}
where $x_v=v/M_\rho$ and ${\cal O}(x^6)$ represents any terms proportional
to $x_v^m x_t^n x_W^p x_Z^q$ such that $m+n+p+q\geq 6$.

When the bottom mass is neglected the $bb$ partial width reads
\begin{equation}\label{eq:GammaBB}
   \Gamma_{\rho^0\rightarrow bb} = \frac{M_{\rho}}{8\pi}
   \left[(g_{\rho bb}^L)^2+(g_{\rho bb}^R)^2\right],
\end{equation}
where
\begin{eqnarray}
   g_{\rho bb}^L &=& \frac{b_L}{4}g'' 
   + \frac{2}{g''}\left[\left(b_L-\frac{4}{3} \right)\frac{M_W^2}{v^2}
   +\frac{1}{3}\;\frac{M_Z^2}{v^2}\right],
   \label{eq:gbbL}\\
   g_{\rho bb}^R &=& p^2\frac{b_R}{4}g'' 
   + \frac{2}{g''}\left(p^2 b_R-\frac{2}{3} \right)
   \left(\frac{M_Z^2}{v^2}-\frac{M_W^2}{v^2}\right).
   \label{eq:gbbR}
\end{eqnarray}
To the leading order in $g''$, the partial width (\ref{eq:GammaBB}) reads
\begin{equation}\label{eq:GammaBBLO}
   \Gamma_{\rho^0\rightarrow bb} = \frac{3}{8}\Gamma_\rho\cdot
   g^{\prime\prime 4} (b_L^2+p^4 b_R^2)\;
   x_v^4.
\end{equation}
Since we ignore the bottom quark mass there are no higher order corrections 
above ${\cal O}(x^4)$ to this expression.

Finally, the $tb$ partial width is 
\begin{equation}\label{eq:GammaTB}
   \Gamma_{\rho^\pm\rightarrow tb} = \frac{M_{\rho}}{8\pi}
   \left[(g_{\rho tb}^L)^2+(g_{\rho tb}^R)^2\right](1-3x_t^2/2+x_t^6/2),
\end{equation}
where
\begin{eqnarray}
   g_{\rho tb}^L &=& -\frac{b_L}{2\sqrt{2}}g''
   +\frac{2\sqrt{2}}{g''}(1-b_L)\frac{M_W^2}{v^2},
   \label{eq:gtbL}\\
   g_{\rho tb}^R &=& -p\frac{b_R}{2\sqrt{2}}g''.
   \label{eq:gtbR}
\end{eqnarray}
To the leading order in $g''$, the partial width (\ref{eq:GammaTB}) reads
\begin{equation}\label{eq:GammaTBLO}
   \Gamma_{\rho^\pm\rightarrow tb} = \frac{3}{4}\Gamma_\rho\cdot
   g^{\prime\prime 4} \left[(b_L^2+p^2 b_R^2)\; x_v^4
   +{\cal O}(x^6)\right].
\end{equation}

\subsection{The branching ratios}
\label{subsec:BRs}

\begin{figure}[ht]
\centerline{
\includegraphics[width=6.5cm]{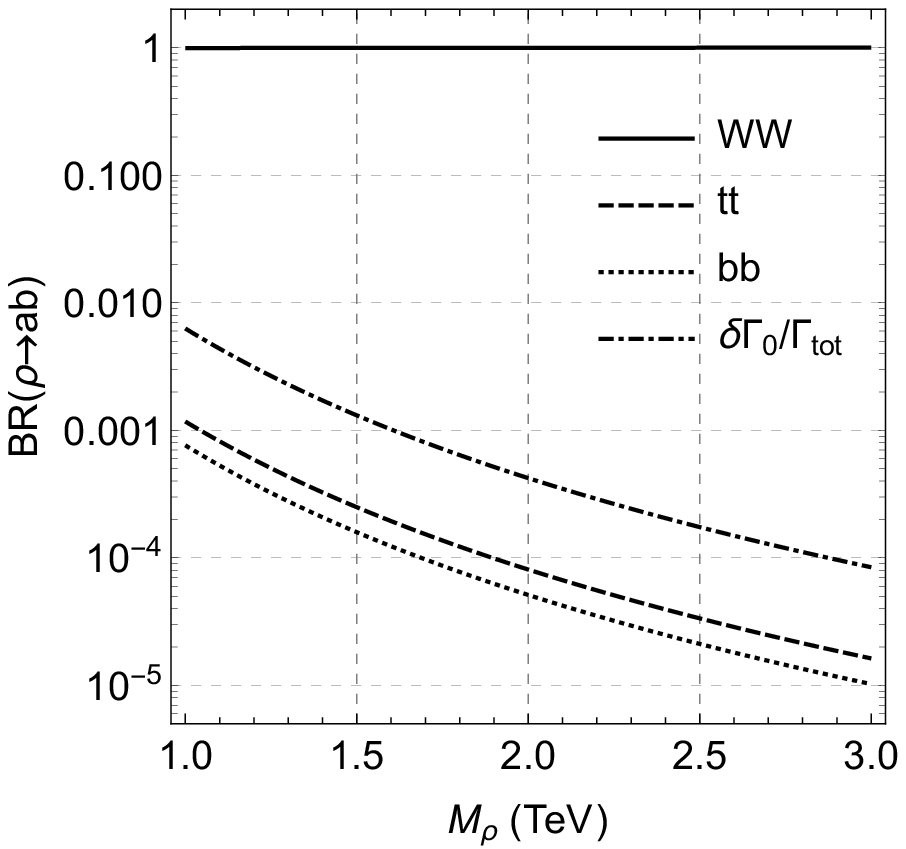}}
\vspace{0.3cm}
\centerline{
\includegraphics[width=6.5cm]{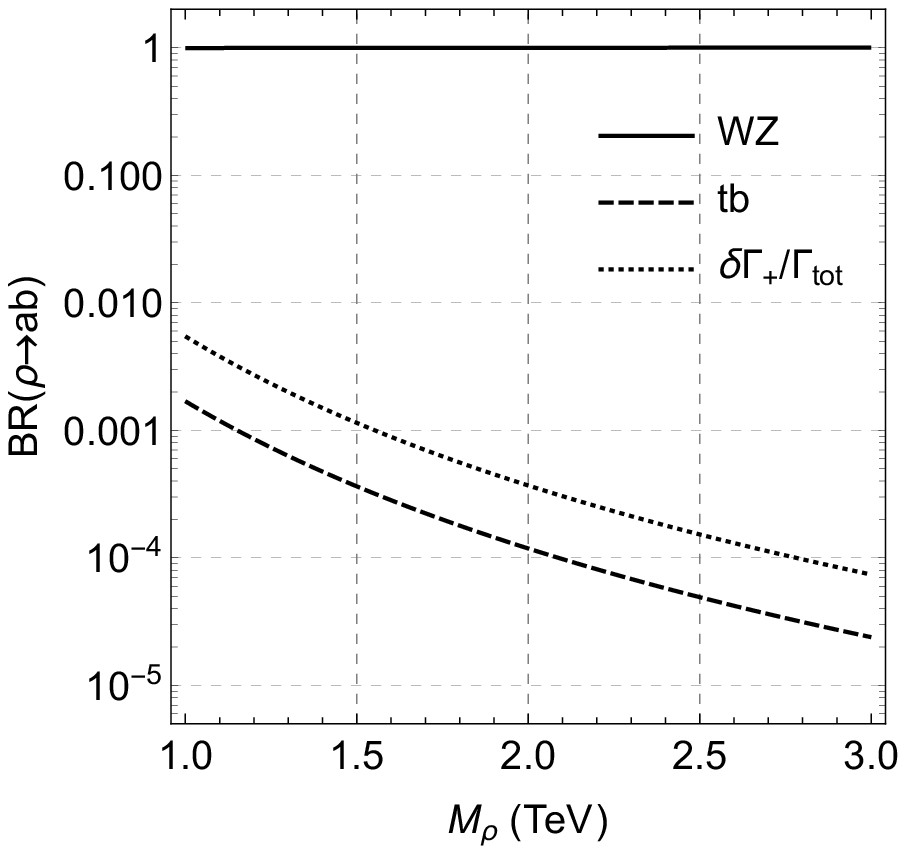}}
\caption{The branching ratios of the neutral (top) and charged (bottom) vector 
resonances as functions of the resonance mass when there are no direct interactions
of the resonance triplet with the top and bottom quarks.}
\label{Fig:BRNDI}
\end{figure}

If the vector resonance triplet does not interact with the fermions directly ($b_{L,R}=0$), 
the branching ratios of all decay channels under consideration do not depend on $g''$. 
This is because all relevant decay widths are proportional to $(1/g'')^2$. 
The no direct interaction branching ratios of the neutral as well as charged vector resonances
are plotted in Fig.~\ref{Fig:BRNDI}. The upper graph depicts 
curves for $\rho^0\rightarrow W^+W^-$, $t\bar{t}$, and $b\bar{b}$. The lower graph 
shows the curves for the $\rho^+\rightarrow W^+Z$ and $t\bar{b}$ channels.
In addition, the upper and lower graphs contain plots of $\delta\Gamma_0/\Gamma_\mathrm{tot}^{(0)}$
and $\delta\Gamma_+/\Gamma_\mathrm{tot}^{(+)}$, respectively, where
\begin{eqnarray}
   \delta\Gamma_0 &=& \Gamma_{ZH}+3\Gamma_{\nu\nu}+3\Gamma_{\ell\ell} 
                       +2\Gamma_{uu} +2\Gamma_{dd},
   \label{eq:deltaGammaN}\\
   \delta\Gamma_+ &=& \Gamma_{WH}+3\Gamma_{\nu\ell}+2\Gamma_{ud},
   \label{eq:deltaGammaP}
\end{eqnarray}
and
\begin{eqnarray}
   \Gamma_\mathrm{tot}^{(0)} &=& \Gamma_{WW}+\Gamma_{tt}(b_L,b_R)
                                           +\Gamma_{bb}(b_L,p^2b_R)+\delta\Gamma_0, 
   \\
   \Gamma_\mathrm{tot}^{(+)} &=& \Gamma_{WZ}+\Gamma_{tb}(b_L,pb_R)+\delta\Gamma_+,
\end{eqnarray}
where we have assumed the same decay widths for the corresponding decay channels
across all generations of leptons as well as across the first two generations of quarks.
The CKM matrix is set to unity throughout the paper.

\begin{figure}[ht]
\centerline{
\includegraphics[width=7.8cm]{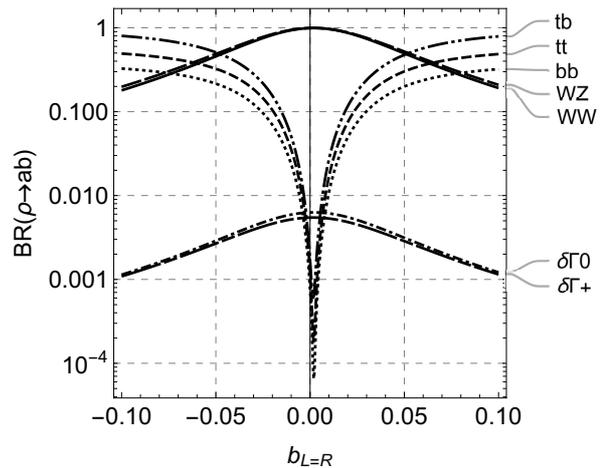}}
\caption{The branching ratios of the neutral and charged vector 
resonances as functions of $b_{L=R}\equiv b_L=b_R$ when $M_\rho=1$~TeV, $g''=18$, and $p=0.75$.
The labels $\delta\Gamma 0$ and $\delta\Gamma +$ indicate the branching ratios corresponding
to the decay widths given in the Eqs.~(\ref{eq:deltaGammaN}) and (\ref{eq:deltaGammaP}),
respectively.}
\label{Fig:BRDI}
\end{figure}

Now, let us consider the situation when the vector resonance triplet interacts directly with
the third quark generation. It means that some or all of the $b_L, b_R, p$ parameters assume
non-zero values. The five parameters of interest that can be varied independently,
$M_\rho$, $g''$, and $b_L, b_R, p$, are too many for displaying
the BR's behavior in a single plot. However, it is instructive to show
BR's for the neutral and charged resonances as functions of their direct interactions to
fermions when $b_L=b_R\equiv b_{L=R}$ and for a specific choice of the values of other parameters.
Namely, we choose $M_\rho=1$~TeV, $g''=18$, and $p=0.75$. The corresponding graph
can be found in Fig.~\ref{Fig:BRDI}.

As expected, the Fig.~\ref{Fig:BRDI} behavior of the branching ratios 
in the vicinity of $b_L=b_R=0$ corresponds well to that observed in Fig.~\ref{Fig:BRNDI}.
When $|b_{L=R}|\lesssim 10^{-2}$ the vector resonance decays are strongly dominated by the 
vector boson channels. The remaining channels contribute less than $1\%$ to the total
decay widths. In addition, in this region, the combined effect of the direct and indirect
interactions of the vector resonance with the top and bottom quarks pulls the branching ratios of
$\rho\rightarrow tt,bb,tb$ even below those of the light fermions. Actually, there are
non-zero values of $b_{L,R}$ for which the direct and indirect interactions cancel each
other out and the vector resonances cease to decay to\footnote{
The effect of the cancellation was studied in detail in our paper~\cite{tBESSprd11}.
We nicknamed the parameter space region where the negative interference of the direct
and indirect interactions suppressed the productions of $t\bar{t}$, $b\bar{b}$, and 
$t\bar{b}/b\bar{t}$
as ``the Death Valley''.}
$tt$, $bb$, and $tb$. 
The glimpse of this effect
can be seen in Fig.~\ref{Fig:BRDI} where the minimum of the $tt,bb,tb$ curves is shifted slightly
to the right of $b_{L=R}=0$, namely at about $b_{L=R}=0.002$.

The branching ratios of the $tt$, $bb$, and $tb$ channels grow fast with the increasing $|b_{L=R}|$. 
It results from the unleashing the large contributions of the first terms
of the couplings~(\ref{eq:gttL}), (\ref{eq:gttR}), 
(\ref{eq:gbbL}), (\ref{eq:gbbR}), (\ref{eq:gtbL}), and (\ref{eq:gtbR}).
These branching ratios reach the branching ratios of the light fermion channels 
at about $|b_{L,R}|\approx 0.01$ and they become comparable with the gauge boson ratios at
about $|b_{L,R}|\approx 0.05$. Of course, the exact numbers depend on the values of 
the model parameters. Fig.~\ref{Fig:BRDI} just illustrates this model's behavior.

\begin{figure}[ht]
\centerline{
\includegraphics[width=8cm]{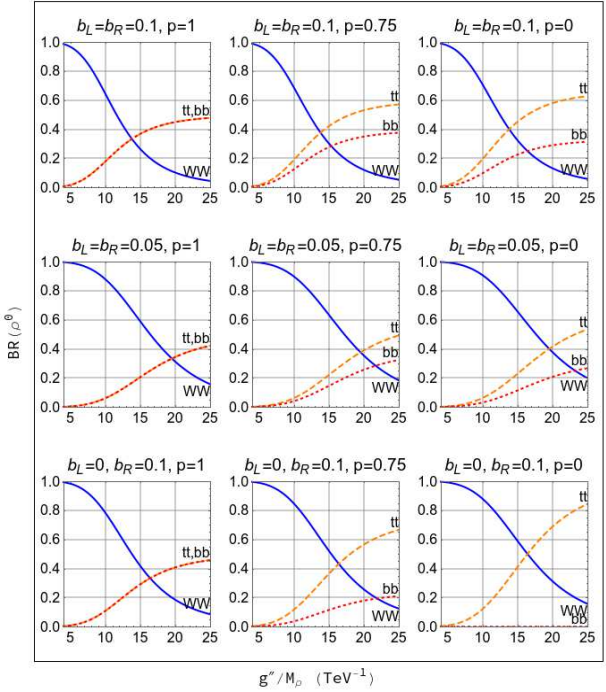}}
\caption{The principal branching ratios of the neutral vector 
resonance as functions of $g''/M_\rho$ in the approximation of 
the leading order in $g''$ and $x_a=M_a/M_\rho=0$, where $a=W,Z,t$.
The blue solid line stands for the $WW$ channel, the orange dashed
line for the $tt$ channel, and the red dotted line represents the $bb$ channel.}
\label{Fig:BRgridN}
\end{figure}

\begin{figure}[ht]
\centerline{
\includegraphics[width=8cm]{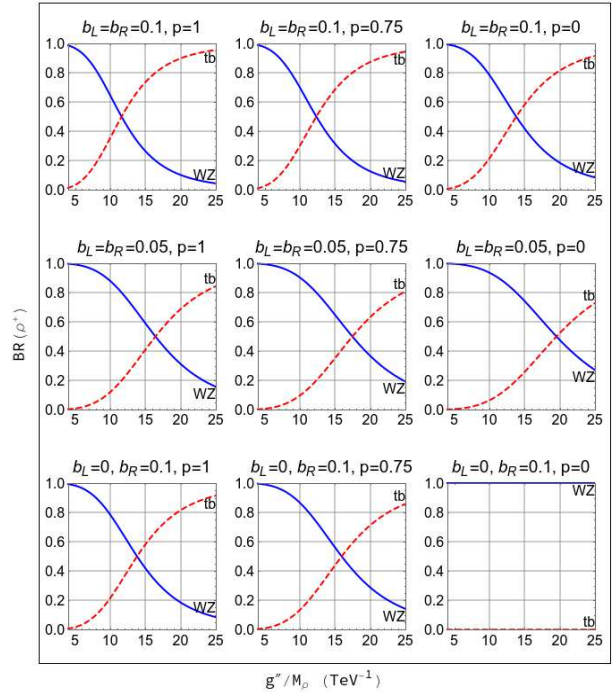}}
\caption{The principal branching ratios of the charged vector 
resonance as functions of $g''/M_\rho$ in the approximation of 
the leading order in $g''$ and $x_a=M_a/M_\rho=0$, where $a=W,Z,t$.
The blue solid line stands for the $WZ$ channel and the red dashed line represents 
the $tb$ channel.}
\label{Fig:BRgridP}
\end{figure}

As the direct interactions grow stronger, the branching ratios of the $tt$, $bb$, and $tb$ channels
become dependent on the $g''/M_\rho$ ratio only. This can be seen in the 
leading-order-in -$g''$ formulas for $tt$, $bb$, and $tb$ channels when $x_a=M_a/M_\rho=0$ for
all relevant final state particles. These approximations introduce deviations from the exact
branching ratios at the level of $x_{W,Z,t}^2$ and higher in the leading-order-in -$g''$ terms.
The branching ratios of the principal decay channels in this approximation are displayed
in Fig.~\ref{Fig:BRgridN} (the neutral resonance) and Fig.~\ref{Fig:BRgridP} (the charged 
resonance).

Nevertheless, the calculations of the mass exclusion limits in this paper
are performed using the exact expressions for the decay widths and branching ratios. 
The approximations discussed above are meant to provide a better insight into the analysis and
a better understanding of the results.

\section{Cross section calculations}
\label{sec:prodXS}

In principle, the current exclusion limits on the masses of new resonances 
result from the comparison of the upper experimental bounds for the resonance 
$s$-channel production cross sections to the model's predictions for this 
observable.  The upper bounds are calculated by the ATLAS and CMS 
Collaborations 
using their data for various final states of various decay channels of the 
vector resonances. Consequently, the provided upper bounds restrict the 
on-shell 
cross sections $\sigma(pp\rightarrow\rho X\rightarrow abX)$. Of course, the 
assumptions and procedures used in the model's prediction calculations have to 
comply with those used by the Collaborations in obtaining the upper bounds.

As long as the decay width of the produced resonance is not too wide for its 
mass,
the cross section $\sigma(pp\rightarrow\rho X\rightarrow abX)$ can be 
conveniently
approximated by the Narrow Width Approximation (NWA) formula
\begin{equation}\label{eq:xsNWAformula}
   \sigma(pp\rightarrow abX)=
   \sigma_\mathrm{prod}(pp\rightarrow \rho X)\times\mathrm{BR}(\rho\rightarrow 
ab),
\end{equation}
where $\sigma_\mathrm{prod}$ is the on-shell cross section for the vector 
resonance production, 
and $\mathrm{BR}(\rho\rightarrow ab)$ is the branching ratio for 
the vector resonance decay channel $\rho\rightarrow ab$.
It is generally expected that the approximation (\ref{eq:xsNWAformula}) works 
well
when $\Gamma_\mathrm{tot}/M_\rho\lesssim 10\%$.
One should also remember that the NWA ignores the signal-background 
interference effects. The influence of these effects on the precision of the 
approximation have been inspected in~\cite{Pappadopulo_etal14}.

The experimental upper bounds have been delivered and updated continually by 
both Collaborations as, over the recent years, the amount of the data collected 
grew and no signs of new particles emerged. The collaborations worked out and 
published the upper bounds not only for the individual decay channels but also 
for their combinations. In addition, for the $WW$ and $WZ$ channels, they were 
able to distinguish between the Drell-Yan (DY) and Vector Boson Fusion (VBF) 
production mechanisms. While, in the former case, the vector resonance is 
produced via the annihilation of the quarks of the colliding protons, in the 
latter, the resonance is created via the fusion of the electroweak vector 
bosons 
emitted by the colliding protons.

In our calculations, $\sigma(pp\rightarrow abX)$ is approximated by the sum of 
the DY and VBF cross sections\footnote{The gluon-gluon production via loops is 
excluded by the Landau-Yang theorem.}. We ignore the top quark involvement in 
the vector resonance production and approximate the CKM matrix by the unity. On 
the other hand, we demonstrate in this Section that the bottom quark 
contribution to the production cannot be neglected. Therefore, the DY 
production of our triplet resonance can proceed only via  $u\bar{u}, d\bar{d}, 
c\bar{c}, s\bar{s}, b\bar{b}\rightarrow \rho^0$, and $u\bar{d}, 
c\bar{s}\rightarrow \rho^+$ (+c.c.).

In the VBF production, the vector resonance can emerge from 
$W^+W^-\rightarrow\rho^0$
and $W^+ Z\rightarrow \rho^+$ (+c.c.). This production is calculated in
the Effective-W Approximation (EWA)~\cite{Dawson1985EWA} considering the longitudinal 
W/Z degrees of freedom only.

\subsection{Production cross section}
\label{subsec:prodXS}

The first factor in the calculation of the cross 
section~(\ref{eq:xsNWAformula}) 
for the given decay channel is the production cross section 
$\sigma_\mathrm{prod}(pp\rightarrow \rho X)$. We address its calculation in 
this Subsection.
The production cross section of a resonance can be expressed as
\begin{equation}\label{eq:prodXS}
   \sigma_\mathrm{prod}(pp\rightarrow \rho X)=\sum_{A\leq B} 16\pi^2 K_{AB}
   F_{AB}\frac{d\Pi_{AB}}{d\hat{s}}|_{\hat{s}=M_\rho^2},
\end{equation}
where $F_{AB}=\Gamma_{\rho\rightarrow AB}/M_\rho$ and $\Gamma_{\rho\rightarrow 
AB}$
are, respectively, the partial fatness and the partial decay width of the 
resonance to 
the partons $A$ and $B$ of the colliding protons. 
Furthermore, $d\Pi_{AB}/d\hat{s}$ is the differential luminosity  of the 
colliding partons, and
\begin{equation}
   K_{AB} = \frac{2J+1}{(2S_A+1)(2S_B+1)}\frac{C}{C_AC_B},
\end{equation}
where $J$ is the spin of the resonance, $C$ is its color factor, and $S_A$, 
$S_B$ and $C_A$, $C_B$ are the spins 
and colors of the initial partons, respectively\footnote{The tBESS vector 
resonance values:
$J_\rho=1$, $C_\rho=1$, $S_q=1/2$, $C_q=3$, $J_{W_L,Z_L}=0$,  $C_{W_L,Z_L}=1$. 
Consequently,
$K_{qq'}=1/12$, $K_{W_LW_L}=K_{W_LZ_L}=3$.}. 
In this paragraph, by ``partons'' we also refer to the electroweak gauge bosons
emitted of the partonic quarks of the colliding protons in the case of the VBF 
production.

Note that the model dependence can
enter the production cross section (\ref{eq:prodXS})
only via the partial decay widths $\Gamma_{\rho\rightarrow AB}$.
The concerned widths include 
$\Gamma_{uu}$, $\Gamma_{dd}$, $\Gamma_{cc}$, $\Gamma_{ss}$, 
$\Gamma_{ud}$, $\Gamma_{cs}$, $\Gamma_{bb}$, $\Gamma_{WW}$ and $\Gamma_{WZ}$ 
where
$\Gamma_{uu}=\Gamma_{cc}$, $\Gamma_{dd}=\Gamma_{ss}$, $\Gamma_{ud}=\Gamma_{cs}$.
In the VBF production processes, all $F_{AB}$'s are proportional to 
$M_\rho^4/g^{\prime\prime 2}$ up to some small corrections of higher order.
All DY partial fatnesses $F_{AB}$, but the $F_{bb}$ one, are proportional
to $1/g^{\prime\prime 2}$ and do not depend on $M_\rho$.
The $bb$ channel is the only one through which the production 
cross section is sensitive to the couplings of the direct interactions. If the 
direct
couplings are sufficiently large then $F_{bb}$ is
virtually proportional to $g^{\prime\prime 2}(b_L^2+p^4b_R^2)$. Neither this 
fatness
depends on $M_\rho$.

The production cross section~(\ref{eq:prodXS}) also depends on the parton 
contents
of the proton via the differential parton luminosities $d\Pi_{AB}/d\hat{s}$,
or ``quasi-luminosities'' for short.
In the DY process, the quasi-luminosities for various partons are obtained from 
the
standard formula
\begin{eqnarray}
       \frac{d\Pi_{AB}}{d\hat{s}}&=&\frac{1}{s}\int_\tau^1 
       \frac{dx}{x}\frac{1}{1+\delta_{AB}}
       \nonumber\\
       &&[f_A(x,\hat{s})f_B(\tau/x,\hat{s})+A\leftrightarrow B], 
\end{eqnarray}
where $s$ and $\hat{s}$ are the squared center of mass energies
of the colliding protons and quarks, respectively, $f_A(x)$ is
a p.d.f.\ of the quark $A$ with the momentum fraction $x$
of its proton's momentum, and $\tau=\hat{s}/s$. The formula for the 
quasi-luminosity
in the VBF production reads
\begin{eqnarray}
   \frac{d\Pi_{AB}}{d\tau}&=&\sum_{i\leq j} \frac{1}{1+\delta_{ij}}
             \int_\tau^1 \frac{dx_1}{x_1} \int_{\tau/x_1}^1 \frac{dx_2}{x_2}
   \nonumber\\
   && [f_i(x_1,q^2) f_j(x_2,q^2) \frac{dL_{A[i]B[j]}}{d\hat{\tau}}
             +i\leftrightarrow j],
\end{eqnarray}
where $\hat{\tau}=\tau/(x_1 x_2)$, and $dL_{A[i]B[j]}/d\hat{\tau}$
is the luminosity for two vector bosons $A$ and $B$ emitted from 
$i$th and $j$th quarks, respectively.

The vector boson luminosity $dL_{A[i]B[j]}/d\hat{\tau}$ is calculated using the 
EWA method. This approach is 
also a subject to some assumptions and restrictions. First of all, the gauge 
bosons are assumed to be emitted on-shell and in small angles to their parental 
quarks. Secondly, the masses of the fusing gauge bosons should be much smaller 
than the produced resonance mass. Finally, the transversal and longitudinal 
polarizations of the emitted gauge bosons are to be considered as separate 
modes.

In the presence of the deviations from the SM, the longitudinal mode usually 
dominates.
Therefore, in our calculations, we consider contributions from this mode only.
The EWA luminosity for two longitudinal vector bosons $A$ and $B$ emitted from 
$i$th and 
$j$th quarks reads
\begin{eqnarray}
   \frac{dL_{A[i]_L B[j]_L}}{d\hat{\tau}}&=&
   \frac{v_{A[i]}^2+a_{A[i]}^2}{4\pi^2} \frac{v_{B[j]}^2+a_{B[j]}^2}{4\pi^2}
   \nonumber\\
   && \frac{1}{\hat{\tau}} [(1+\hat{\tau})\log(1/\hat{\tau})-2(1-\hat{\tau})],
\end{eqnarray}
where $v_{A[i]}$ and $a_{A[i]}$ are the vector and axial-vector couplings of 
the 
electroweak gauge boson $A$ to the quark current of $q_i$, respectively.

\begin{figure}[ht]
\centerline{
\includegraphics[width=8cm]{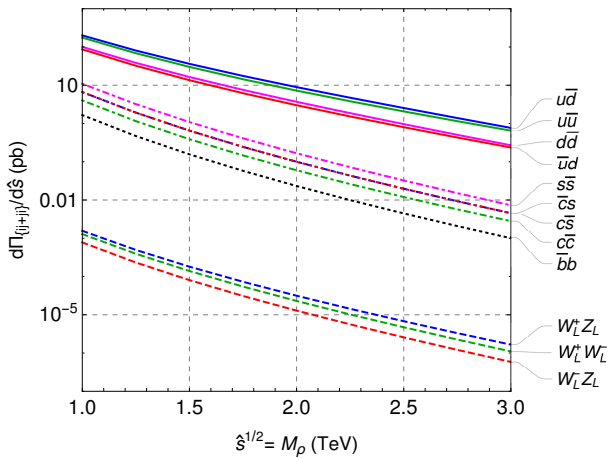}}
\caption{Quasi-luminosities for the DY and VBF LHC processes at 
$\sqrt{s}=13$~TeV
as functions of the CMS energy of the colliding partons.}
\label{Fig:PseudoLum}
\end{figure}

In Fig.~\ref{Fig:PseudoLum}, the quasi-luminosities for the DY and VBF LHC 
processes at 
$\sqrt{s}=13$~TeV are shown. For the numerical evaluation the \verb+Mathematica+
package \verb+ManeParse+ \cite{ManeParse2016} with the CT10 p.d.f.\ set from 
the LHAPDF 6 library in 
the HepForge repository~\cite{HepForgeCT10_2015} is used. 
As can be seen, the DY quasi-luminosities dominate by many orders of magnitude 
over the quasi-luminosities of the VBF production. It is because the VBF 
production process is suppressed against the DY production process by two 
orders in the perturbative expansion. 

The relative sizes of the individual DY quasi-luminosities can be understood in 
terms of the proton parton contents for individual flavors. The valence quark 
quasi-luminosities clearly stand over the sea quark ones. The $bb$ 
quasi-luminosity is the smallest one by 2 -- 3 orders of magnitude below the $u$ 
and $d$ related quasi-luminosities. This poses a question whether the 
contribution of the vector resonance production via the sea quark annihilation 
are worthy of consideration. Of these, the $b\bar{b}\rightarrow\rho^0$ 
production is the most disputable. On the one hand, the $b\bar{b}$ 
quasi-luminosity contributes the least. On the other hand, this is the only 
production process that is sensitive to the direct interactions.

\begin{figure}[ht]
\centerline{
\includegraphics[width=8cm]{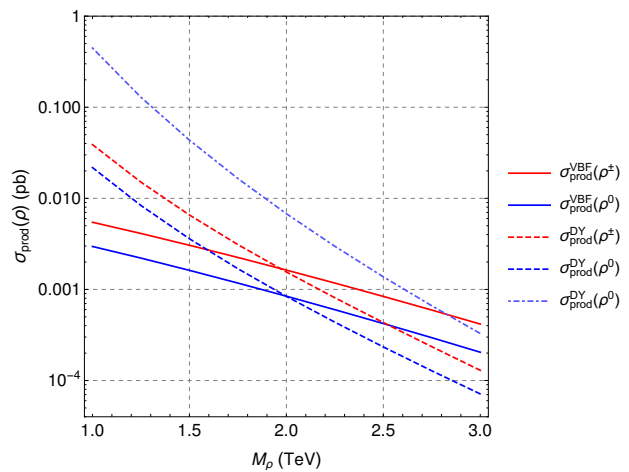}}
\caption{The production cross sections $\sigma_\mathrm{prod}(pp\rightarrow\rho 
X)$ for the DY and VBF productions of the neutral (blue) and charged (red) 
vector resonances as functions of the resonance mass when $g''=20$. The solid lines 
stand for the VBF production which is insensitive to the direct interactions. 
The dashed lines denote the DY production without the direct interactions. The 
dot-dashed line shows the neutral DY case with the direct interactions, namely 
$b_L=b_R=0.1$, $p=1$.}
\label{Fig:ProdXS}
\end{figure}

To address this issue the production cross section~(\ref{eq:prodXS}) is plotted 
as a function of the resonance mass for the DY and VBF modes assuming $g''=20$. 
In the DY case, we also distinguish between the situations with and without the 
direct interactions. However, only the neutral resonance DY production is 
sensitive to the direct interactions. Therefore, turning the direct interactions 
on by setting $b_L=b_R=0.1$, $p=1$ affects only the neutral DY mode. All these 
plots are shown in Fig.~\ref{Fig:ProdXS}. The plotted production cross sections 
are calculated considering contributions of all quark flavors but the top quark 
to the proton partonic contents.

The following observations can be made in Fig.~\ref{Fig:ProdXS}. 
First of all, in spite of much smaller 
quasi-luminosity, the VBF production cross section is comparable with the DY one.
It is because the VBF production quasi-luminosity handicap is counter-balanced
with the dominance of $F_{WW,WZ}$ over $F_{qq}$ (see Figs.~\ref{Fig:BRNDI} and 
\ref{Fig:BRDI}).
Therefore, both production modes must be considered in our analysis.
The different slopes of the DY and VBF curves can be understood
when we recall that $F_{qq}\propto M_\rho^0$ while $F_{WW,WZ}\propto M_\rho^4$.

Secondly, we can see that the direct interactions can have a visible impact on 
the production cross section for the neutral DY mode. When $M_\rho=1$~TeV and $g''=20$ 
the contributions of $b\bar{b}\rightarrow\rho^0$
to the production cross section is $12\%$ and $95\%$ when $b_L=b_R=0.01$, $p=1$ 
and $b_L=b_R=0.1$, $p=1$, respectively.

Since the sea-without-$b$ quark production of the vector triplet is not 
sensitive to the direct interactions,
it is reasonable to expect that its contribution to the production cross 
section will be small under all circumstances. The size of this contribution cannot be read off of 
Fig.~\ref{Fig:ProdXS}. In our calculations, we have established that
by ignoring the sea quark production in the no-direct-interaction situation
the production cross section is lowered by about $7\%$ when $M_\rho=1\unit{TeV}$
and by about $5\%$ at $M_\rho=3\unit{TeV}$. The discrepancy decreases as $M_\rho$ grows.

\subsection{$\sigma_\mathrm{prod}\times \mathrm{BR}$}
\label{subsec:XSprodTimesBR}

Following Eq.~(\ref{eq:xsNWAformula}), $\sigma(pp\rightarrow \rho X\rightarrow 
abX)$ is obtained when the production cross section is multiplied by the 
branching ratio of $\rho\rightarrow ab$. When there are no direct interactions 
of the vector triplet with top and bottom quarks more than $99\%$ of $\rho$ 
decays into the $WW$/$WZ$ channel. It can be seen in Fig~\ref{Fig:BRNDI}. To a 
high accuracy, this assertion holds for any considered values of $g''$ and 
$M_\rho$. Therefore, $\sigma_\mathrm{prod}\times \mathrm{BR}(\rho\rightarrow 
WW/WZ)$ is virtually identical with the production cross section 
$\sigma_\mathrm{prod}$.

The direct interactions influence the resulting cross sections solely via the 
branching ratios with a single exception. The exception is the neutral resonance 
production via the DY mode. There, the production depends on the parameters of 
the direct interactions as well. Recall that the dependence originate in the 
$b\bar{b}\rightarrow \rho^0$ vertex.

The direct interactions can alter the vector resonance branching ratios of the 
individual channels significantly. It can be inferred from the behavior of 
$\mathrm{BR}(\rho\rightarrow WW/WZ)$ and $\mathrm{BR}(\rho\rightarrow tt/bb/tb)$ 
as it is depicted in Figs.~\ref{Fig:BRDI}, \ref{Fig:BRgridN}, and 
\ref{Fig:BRgridP}. The prevailing behavior of these branching ratios is that 
$\sigma_\mathrm{prod}\times \mathrm{BR}(\rho\rightarrow tt/bb/tb)$ grows with the 
strength of the direct interactions while $\sigma_\mathrm{prod}\times 
\mathrm{BR}(\rho\rightarrow WW/WZ)$ decreases. Nevertheless, there are 
particular combinations of the values of $b_{L,R}$ and $p$ when this statement 
does not hold. For example, if $b_L=0$ and $p=0$ the vector resonance decays 
strongly to $tt$ while it decays to $bb$ only via the indirect interactions not 
sensitive to these parameters. In such a case, $\mathrm{BR}(\rho\rightarrow tt)$ 
grows with $b_R$ while $\mathrm{BR}(\rho\rightarrow bb)$ decreases.
This very behavior will transfer without alteration into the cross sections of 
the processes where the neutral resonance is not DY produced. Otherwise, the 
cross section dependence on the direct interaction parameters will result from 
the interplay between the BR and production cross section dependences.

In our previous paper~\cite{tBESSapp18}, we investigated the tBESS mass 
exclusion limits for the no direct interactions case. In this analysis, all sea 
partonic quarks, including the $b$ quarks, were ignored. It was a well-justified 
assumption for the case. Nevertheless, in the paper, we also commented on our 
expectation regarding the exclusion limits for the case with the direct 
interactions. Neglecting the $b$ quarks, we expected that the direct 
interactions would lower the cross sections for the $WW$/$WZ$ channels and 
increase the cross sections for the $tt,bb$/$tb$ channels. Since the latter 
channels did not reach the experimental upper limits for any admissible values 
of $b_{L,R}$, $p$, we predicted that the presence of the direct interactions 
would relax the mass exclusion limits. Now, we understand that the bottom quark 
contribution to the vector resonance production can be ignored only for 
particular selections of the direct interaction parameters. Thus, our expectation
was not correct except for these particular cases.

\section{The exclusion limits on the vector resonance mass}
\label{sec:masslimits}
In this Section, we work out the mass exclusion limits for the vector triplet
of our model by the confrontation of its cross sections~(\ref{eq:xsNWAformula})
with the experimental upper bounds provided by the ATLAS and CMS Collaborations.
The regions of the parameter space where the predicted cross section exceeds
the upper bounds are excluded.
We review fourteen vector resonance decay channels available to the date
of this analysis:
$WZ$~\cite{P5,N1,44,M7},
$WW$~\cite{P5,43,N1},
$WH$~\cite{P5,M4}, 
$ZH$~\cite{P5,M4}, 
$jj$~\cite{N9,N10}, 
$\ell\ell$~\cite{N2}, 
$\ell\nu$~\cite{N6}, 
$\tau\tau$~\cite{M3}, 
$\tau\nu$~\cite{M2}, 
$tt$~\cite{R10},
$bb$~\cite{P2},
and 
$tb$~\cite{P4,N3}, where $\ell=e,\mu$. 
The corresponding $95\%$~C.L.\ bounds used in this Section
are based on the integrated luminosity of about $36\unit{fb}^{-1}$ (full 2016 
data) or less.
As was discussed in previous Sections, various experimental and theoretical 
considerations
restrict our mass exclusion limit searches to the following region of the 
parameter space: 
$1\leq M_\rho/\mathrm{TeV}\leq 3$, $12\leq g''\leq 25$ 
$|b_{L,R}|\leq 0.1$, $0\leq p \leq 1$.

As $g''$ approaches the naive perturbative limit of $25$, the radiative 
corrections grow and our tree-level results become less 
reliable. However, various precision levels are needed for addressing various questions. 
Therefore, we present results for $g''$ up to the naive perturbative limit 
and leave up to the reader to choose the maximal trustworthy value of $g''$
for his/her purpose. In addition,
while the inclusion of the radiative corrections would alter the obtained exclusion 
limits it does not invalidate the qualitative lessons learned from our analysis.

After the analysis advertised above has been finalized new experimental upper 
bounds
have been published by the ATLAS and CMS Collaborations. Some of them were 
still based on
the $36\unit{fb}^{-1}$~\cite{R6,R9}
dataset, other stemmed from the bigger 
$77-80\unit{fb}^{-1}$~\cite{R5,R8} and
$139\unit{fb}^{-1}$~\cite{R2,R3,R4} datasets. 
At the end of this Section we append an update on
how these new upper bound affect our mass exclusion limits. Nevertheless, 
since the analysis of all collected LHC data
by the Collaborations is still in progress, even these updates of the mass 
limits can become
obsolete in the near future. Yet, we believe that the presented work provide 
valuable
experience independent of the actual values of the mass exclusion limits.

\subsection{No direct interactions}

The no direct interactions case was already analyzed in our previous 
work~\cite{tBESSapp18}. Out of all inspected channels there, only the $WW$ and 
$WZ$ channels provided the exclusion limits for the vector resonance mass. 

In this paper, the mass exclusion limits based on the updated upper bounds are 
presented. Besides the channels considered in~\cite{tBESSapp18}, a new decay 
channel, the $\tau\nu$ one \cite{R6}, has also been added to the current study. 
The distinction of the DY and VBF production modes for the $WW/WZ$ upper bounds 
is 
another novelty. Finally, the proton partonic contents includes the $s$, $c$, 
and $b$ quarks in the calculation of the model's cross section. All these 
updates have not altered the conclusion of the previous paper~\cite{tBESSapp18} 
that the $WW$ and $WZ$ channels are the only channels providing the mass 
exclusion limits for the tBESS vector triplet with no direct interactions. Of 
course, the limits themselves have been changed by this analysis upgrade.

\begin{figure*}[thb]
\centerline{
\includegraphics[width=17cm]{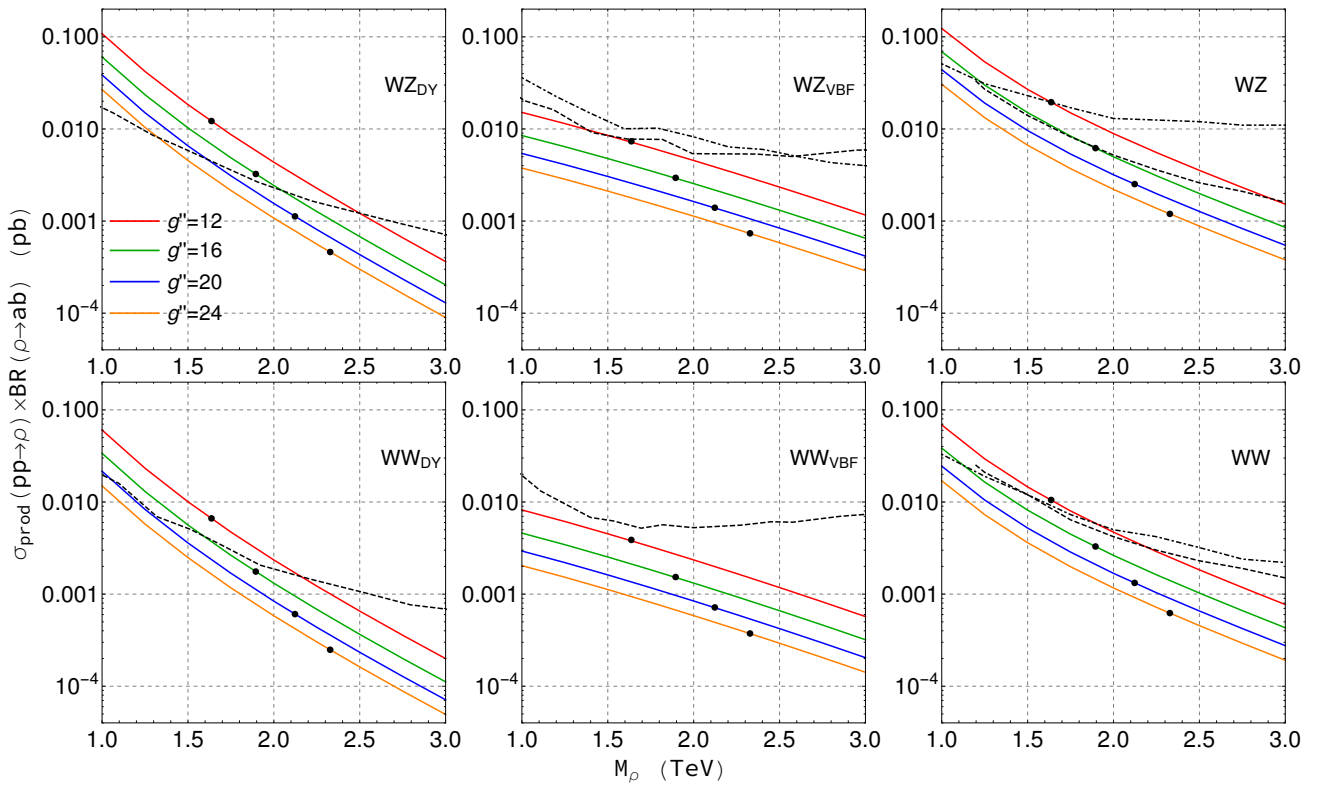}}
\caption{The plots of $\sigma_\mathrm{prod}\times\mathrm{BR}$ predicted 
by the tBESS model
for the $WW$ and $WZ$ channels (solid lines) along with the relevant 
experimental
upper bounds (dashed and dot-dashed lines) as functions of $M_\rho$ assuming no 
direct
interactions of the vector triplet with the top and bottom quarks. The predicted
cross sections are shown for four different values of $g''$: $12$ (red), $16$ 
(green), 
$20$ (blue), $24$ (orange). They decrease with $g''$. The black dots
indicate the resonance mass at which the resonance fatness amounts to $10\%$.
The first row of the graphs
correspond to the $WZ$ final state, the second one to the $WW$ final state. The 
first, second
and third columns of the graphs correspond to the DY production, the VBF 
production and the
combination of both, respectively. In all graphs, the 13~TeV proton-proton 
collisions are 
considered.}
\label{Fig:MELNDI}
\end{figure*}

Taking into account the separate upper bounds for both production modes of the 
$WW$ and $WZ$ channels, our exclusion limits are based on the following six 
processes: a) the DY production of $WW$ and $WZ$, b) the VBF 
production\footnote{
The sensitivity of the LHC vector boson scattering processes,
$WZ\rightarrow WZ$ and $WW\rightarrow WW$, to new vector resonances
was also investigated recently in~\cite{Delgado}.
}
of $WW$ and $WZ$, and c) the DY+VBF production of $WW$ and $WZ$. In 
Fig.~\ref{Fig:MELNDI}, the tBESS cross sections for all these modes as 
functions 
of $M_\rho$ are shown at $g''=12$, $16$, $20$, and $24$. In addition, the 
applicable experimental upper bounds from the six channels mentioned above are 
superimposed on the graphs. The mass regions where the predicted cross section 
exceeds the experimental upper bound are experimentally excluded.

To avoid the cluttering of the Fig.~\ref{Fig:MELNDI} graphs with unnecessary 
curves, only the most restricting experimental upper bounds are displayed 
there. In 
the $WZ_\mathrm{DY}$, $WW_\mathrm{DY}$, and $WW_\mathrm{VBF}$ graphs, the most 
restricting bounds are provided by a single curve. In the remaining cases, the 
most restrictive upper bounds are comprised of two curves, each providing the 
most restrictive bound for a different subregion of the considered resonance 
mass interval of 1 -- 3~TeV. The sources of the displayed upper bound curves of 
the individual channels are summarized in Table~\ref{Tab:MELNDIsources}. The 
table also indicates the particular final states that were used to obtain the 
bounds. 

\begin{table}[htb]
\centering
\caption{The sources of the experimental upper bounds for the $WW$ and $WZ$ 
channels 
depicted in Fig~\ref{Fig:MELNDI}.
Beside the decay channel, the first column also indicates whether the 
considered production proceeds
via the DY production, the VBF production or both. The second column shows what 
is the 
integrated luminosity of the data sample used to derive the bound.
The third column indicates the final states through which the given channel was 
observed.
The bounds were provided by the ATLAS and CMS Collaborations in the papers 
referred to in 
the last column.}
\label{Tab:MELNDIsources}
\begin{tabular}{lccc}
\hline
channel & luminosity & final state & reference\\
 & $\mathrm{fb}^{-1}$ & & \\
\hline
$WZ_\mathrm{DY}$ & $36.1$ & $qqqq+\nu\nu qq+\ell\nu qq$&
\cite{P5}\\
&& $+\ell\ell qq+\ell\ell\ell\nu$ & 
\\
$WZ_\mathrm{VBF}$ & $36.1$ & $\nu\nu qq+\ell\nu qq+\ell\ell qq$ 
&\cite{P5}\\
&&$+\ell\ell\ell\nu$&\\
& 36.1 &$\ell\ell qq+\nu\nu qq$&\cite{M7}
\\
$WZ$ & $35.9$ & $jj$ &\cite{N1}\\
&$13.2$&$\ell\ell qq$&\cite{44}
\\
$WW_{DY}$ & $36.1$ & $qqqq+\ell\nu qq+\ell\nu\ell\nu$ &\cite{P5}
\\
$WW_{VBF}$ & $36.1$ & $\ell\nu qq+\ell\nu\ell\nu$ & \cite{P5}
\\
$WW$ & $35.9$ & $jj$ &\cite{N1} \\
& $13.2$ & $\ell\nu qq$ & \cite{43} \\
\hline
\end{tabular}
\end{table}

In the tBESS model, the neutral and charged vector resonances are 
virtually degenerate in mass. Therefore, the exclusion limit is obtained as the 
higher one 
of the charge and neutral limits.
In particular, we take the most stringent of the limits found in all six 
processes displayed in 
Fig.~\ref{Fig:MELNDI}. 
Of course, we can also establish the exclusion limits for individual charge 
modes 
independently of each other. The lessons learned from the tBESS vector
triplet model can be applied to other models with similar phenomenological 
traits.

The tBESS exclusion mass limits for different values of $g''$ within $14\leq 
g''\leq 25$ are listed in Table~\ref{Tab:MELNDI}. The exclusion limits for 
$g''$ 
values below $14$ are not shown because their fatness exceeds $40\%$ which 
makes the limits obtained via the NWA calculations unreliable.

\begin{table*}[htb]
\centering
\caption{The mass exclusion limits (MEL) for the tBESS vector resonance triplet 
without direct 
interactions for various values of $g''$. The second row contains the values of 
the resonance
fatness. All MEL values shown in the table indicate the upper boundaries of the 
mass exclusion
region.}
\label{Tab:MELNDI}
\begin{tabular}{c|cccccccccccc}
\hline
$g''$ & 14 & 15 & 16 & 17 & 18 & 19 & 20 & 21 & 22 & 23 & 24 & 25\\
\hline
$\Gamma_\mathrm{tot}/M_\rho$ 
& 0.39 & 0.21 & 0.14 & 0.10 & 0.07 & 0.05 & 0.03 & 0.02 & 0.02 & 0.01 & 0.01 & 
$<0.01$ \\
MEL (TeV) & 2.50 & 2.21 & 2.07 & 1.95 & 1.83 & 1.70 & 1.60 & 1.51 & 1.43 & 1.37 
& 1.31 & 1.24\\
\hline
\end{tabular}
\end{table*}

\subsection{The direct interactions included}

In this paper, the impact of the direct interactions on the tBESS vector 
triplet 
mass exclusion limits is studied for the first time. The direct interactions 
can 
affect the limits by increasing the cross sections of the top and bottom quark 
decay channel processes. It remains to be seen if the increase is sufficient 
for 
the cross sections to reach the experimental upper bounds, in some regions of 
the parameter space at least. Certainly, the direct interactions also affect 
the 
cross sections of the $W$ and $Z$ decay channel processes. Thus, the mass 
exclusion limits will be influenced by the direct interactions even if the top 
and bottom quark processes fail in providing additional restrictions.

The current upper bounds based on the data from the remaining channels --- with 
leptons, light quarks, and the Higgs boson --- are too weak to modify the tBESS 
triplet mass exclusion limits. Even though the introduction of the direct 
couplings does affect the tBESS cross sections of these channels, their values 
remain far below their upper bounds. They also cannot compete with the 
$tt/bb/tb$ cross sections once the strength of the direct interactions exceeds 
the level of $|b_{L,R}|\geq 0.02-0.03$. Therefore these channels are not able 
to 
contribute to the tBESS mass exclusion limits at the current amounts of the 
collected data.  We are not going to further discuss these channels in this 
paper.

\subsubsection*{$tb$, $tt$, $bb$}

\begin{figure}[htb]
\centerline{
\includegraphics[width=8cm]{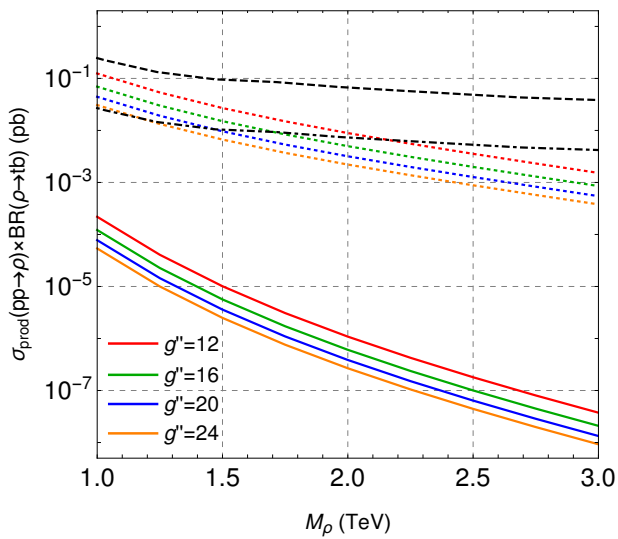}}
\caption{The plots of $\sigma_\mathrm{prod}\times\mathrm{BR}$ predicted 
by the tBESS model
for the $tb$ channel when there are no direct interactions (colored solid 
lines) 
and when $\mathrm{BR}(\rho\rightarrow tb)=1$ (colored dotted lines) 
along with the relevant experimental upper bounds (black dashed line) as 
functions of $M_\rho$.
The black dot-dashed line depicts the expected upper bound for the integrated 
luminosity 
$3000\unit{fb}^{-1}$.
The predicted cross sections are shown for four different values of $g''$: $12$ 
(red), $16$ (green), 
$20$ (blue), $24$ (orange). They decrease with $g''$. The $13\unit{TeV}$ 
collisions
of protons are assumed.}
\label{Fig:MELDItb}
\end{figure}

The $tb$ channel cross section does not provide the mass exclusion limits for 
any considered values of the model parameters. In this channel, the sensitivity 
to the direct interactions enters via the $\mathrm{BR}(\rho\rightarrow tb)$. 
With the growing strength of the direct interactions, the branching ratio also 
grows, reaching the limiting value of $1$ as 
$(b_L^2+p b_R^2)\rightarrow\infty$. Note, however, that there are regions of 
the parameter space where $\mathrm{BR}(\rho\rightarrow tb)$ exceeds $90\%$ 
already at $|b_{L,R}|=0.1$ (see Figs.~\ref{Fig:BRDI} and \ref{Fig:BRgridP}). In 
principle, $\sigma_\mathrm{prod}\times\mathrm{BR}(\rho\rightarrow tb)$ can 
assume any value between the no direct interaction cross section and the 
production cross section if the suppressing Death Valley region of the 
parameter 
space is ignored. In Fig.~\ref{Fig:MELDItb}, the plots of these two extremes of 
the tBESS cross sections as functions of $M_\rho$ for various $g''$ are shown 
along with the most restricting experimental upper bounds for this channel. 
It demonstrates that the current experimental bounds do not exclude the 
production 
cross section values. In 
addition, the estimated upper bound, when the integrated luminosity reaches 
$3000\unit{fb}^{-1}$ (HL-LHC), is also plotted in the graph\footnote{ The 
hypothetical HL-LHC bound is obtained by a simple rescaling of the current 
$35.9\unit{fb}^{-1}$ bound by the factor of $\sqrt{35.9/3000}$.}. 
It seems that the $3000\unit{fb}^{-1}$ luminosity will be needed to 
restrict the vector resonance mass from the data in this channel.

As far as the $tt$ and $bb$ channels are concerned, they are sensitive to the 
direct interactions through the $b\bar{b}\rightarrow\rho^0$ production, as well 
as the vector resonance decays to $t\bar{t}$ and $b\bar{b}$. Nevertheless, in 
these channels, no mass exclusion limits for the parameter space region under 
consideration are implied by the current experimental data. The list of the 
currently most restricting upper bounds, this conclusion is based upon, is 
shown 
in Table~\ref{Tab:MELDItbsources}.

\begin{table}[htb]
\centering
\caption{The sources of the experimental upper bounds for the $tt$, $bb$, and 
$tb$ channels
we use in our analysis.}
\label{Tab:MELDItbsources}
\begin{tabular}{lccc}
\hline
channel & luminosity & final state & reference\\
\hline
$tt$ & 35.9$\unit{fb}^{-1}$ & dileptons+lepton & \cite{R10}
\\
& & +hadronic &
\\
$bb$ & 36.1$\unit{fb}^{-1}$ & 2-$b$ jets & \cite{P2}
\\
$tb$ & 36.1$\unit{fb}^{-1}$ & semileptonic & \cite{P4}
\\
& & +hadronic &
\\
\hline
\end{tabular}
\end{table}

\subsubsection*{$WW$, $WZ$}

In the no direct interaction case, the $WW$ and $WZ$ channels provided the only 
mass exclusion limits for the tBESS vector triplet. Once the direct 
interactions 
are introduced, the predicted cross section gets modified and become dependent 
on the values of $b_{L,R}$ and $p$. The $WZ_\mathrm{DY}$, $WZ_\mathrm{VBF}$, 
$WZ_\mathrm{DY+VBF}$, and $WW_\mathrm{VBF}$ cross sections diminish in 
comparison with the no direct interaction case for all parameter values under 
consideration. It is because these cross sections depend on $b_{L,R}$ and $p$ 
through $\mathrm{BR}(\rho\rightarrow WW)$ and $\mathrm{BR}(\rho\rightarrow WZ)$ 
only.

The behavior of the $WW_\mathrm{DY}$ cross section is not so easy to 
conjecture. 
Its sensitivity to the direct interactions originates not only in 
$\mathrm{BR}(\rho\rightarrow WW)$ but also in the production of $\rho$ through 
the $b\bar{b}$ annihilation. Of course, this feature impacts the exclusion 
limits obtained from the combined (DY+VBF) $WW$ cross section.

When $|b_{L,R}|\leq 0.1$ the cross sections for all three $WZ$ modes
lie in the bands between the no direct interaction cross section (the upper 
boundary) and
the $b_L=-b_R=-0.1$ and $p=1$ cross section (the lower boundary)\footnote{
The $WZ_\mathrm{DY,VBF,DY+VBF}$ and $WW_\mathrm{VBF}$ cross sections for all 
combinations of 
$b_{L,R}$ values such that $|b_{L,R}|=0.1$ and $p=1$ are virtually identical.}.
The upper boundary of the cross section stripe for the $WW_\mathrm{VBF}$ mode 
is also determined 
by the no direct interaction cross section. The lower boundary is given by the 
cross section at
$b_L=b_R=-0.1$ and $p=1$.

\begin{figure*}[thb]
\centerline{
\includegraphics[width=17cm]{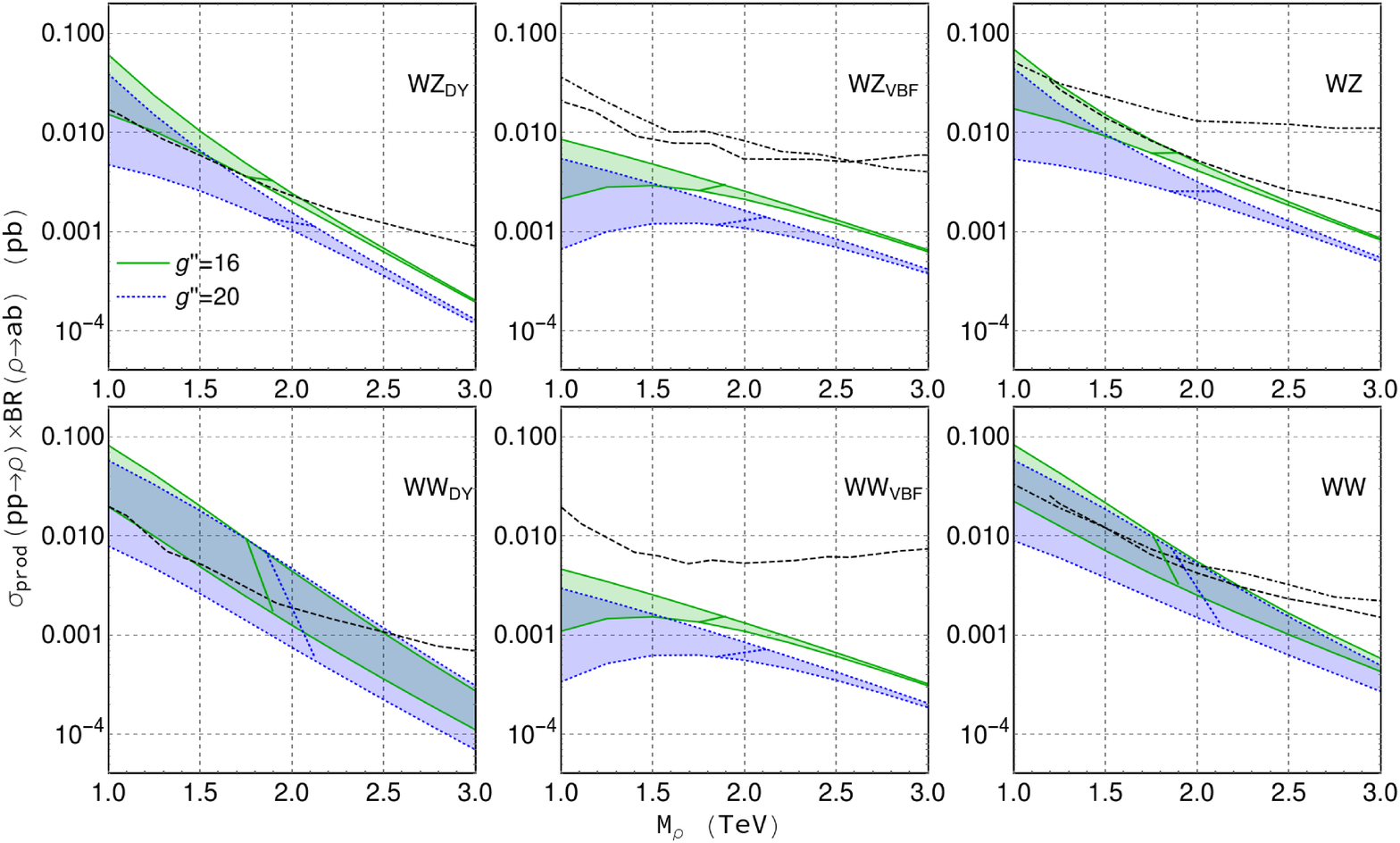}}
\caption{The bands of possible values of 
$\sigma_\mathrm{prod}\times\mathrm{BR}$ predicted 
by the tBESS model for the $WW$ and $WZ$ channels produced via the DY, VBF and
DY+VBF production modes. The bands are constructed for $g''=16$ (solid green) 
and
$g''=20$ (dotted blue) when the direct interaction parameters are restricted to
$|b_{L,R}|\leq 0.1$ and $0\leq p \leq 1$. 
The bands are crossed by the lines of the same style as the bands' boundaries
indicating the resonance mass at which the resonance fatness amounts to $10\%$.
The same experimental
upper bounds (dashed and dot-dashed black lines) as in Fig.~\ref{Fig:MELNDI} 
are also shown.
In all graphs, the 13~TeV proton-proton collisions are considered.}
\label{Fig:MELDIWZ}
\end{figure*}

Regarding the $WW_\mathrm{DY}$ and $WW_\mathrm{DY+VBF}$ modes, their cross 
sections can either grow or decrease with the growing strength of the direct 
interactions, depending on the particular combination of the $(b_{L,R},p)$ 
parameter values. This more complex behavior originates from the competition 
between the growing production cross section and the shrinking branching ratio. 
We determined numerically that when $|b_{L,R}|\leq 0.1$ the $WW_\mathrm{DY}$ 
and 
$WW_\mathrm{DY+VBF}$ cross sections are bound from below by the cross section 
for $b_L=0, b_R=-0.1, p=0$. From above, they are bound by the $b_L=-b_R=-0.1$ 
and $p=1$ cross section.
This behavior is illustrated in Fig.~\ref{Fig:MELDIWZ} where the bands of 
possible values of the $WW$ and $WZ$ cross sections as functions of $M_\rho$, 
when $|b_{L,R}|\leq 0.1$ and $0\leq p \leq 1$, are depicted. The bands are 
constructed for two values of $g''$, namely $16$ and $20$. In addition, the 
same 
experimental upper bounds, as in Fig.~\ref{Fig:MELNDI}, are superimposed. We 
can 
see that the $WW_\mathrm{VBF}$ and $WZ_\mathrm{VBF}$ provide no mass exclusion 
limits for the vector triplet of our model, while the remaining modes do. 

Since there are too many free parameters involved, and the behavior of some 
modes is not simple, it is not possible to find a way to present the mass 
exclusion limits in a concise manner. Nevertheless, the graphs in 
Fig.~\ref{Fig:MELDIWZ} provide information about certain aspects of the mass 
exclusion limit behavior: whether there are any limits at all, what is the 
range 
of their possible values, and so on. 

If additional restrictions on the free parameters were imposed more specific 
information about the exclusion limits could be obtained. For example, let us 
reduce the number of free parameters of the tBESS model by assuming that 
$b_L=b_R$ and $p=1$. It means that the direct interactions are parameterized by 
a single free parameter $b\equiv b_L=b_R$. In Tab.~\ref{Tab:MELDI1}, the 
smallest mass exclusion limits that can be found from the current $WW$ and $WZ$ 
experimental upper bounds for various values of $g''$ when $|b|\leq 0.1$ are 
shown. The table also displays the value of $b$ and the value of the resonance 
fatness $\Gamma_\mathrm{tot}/M_\rho$ that corresponds to the found mass 
exclusion limit.

\begin{table*}[htb]
\centering
\caption{The smallest tBESS resonance mass exclusion limits (MEL)
within the interval $|b_{L=R}|\leq 1$ assuming $b_{L=R}\equiv b_L=b_R$ and $p=1$
for various values of $g''$.
The second and third rows contain, respectively, the values of $b_{L=R}$ and 
of the resonance fatness $\Gamma_\mathrm{tot}/M_\rho$
that correspond to the quoted mass exclusion limit. 
The excluded masses lie below MEL.}
\label{Tab:MELDI1}
\begin{tabular}{c|cccccccccccccc}
\hline
$g''$ & 14 & 15 & 16 & 17 & 18 & 19 & 20 & 21 & 22 & 23 & 24 & 25\\
\hline
$b_{L=R}\times 10^2$ & 9.5 & 5.6 & 4.4 & 3.6 & 3.2 & 2.8 & 2.5 & 2.0 & 1.7 & 
1.7 & 1.7 & 1.6 \\
$\Gamma_\mathrm{tot}/M_\rho$ 
& 0.36 & 0.20 & 0.14 & 0.10 & 0.06 & 0.04 & 0.03 & 0.02 & 0.02 & 0.01 & $\leq 
0.01$ & $\leq 0.01$ \\
MEL (TeV) & 2.42 & 2.15 & 2.04 & 1.92 & 1.77 & 1.64 & 1.53 & 1.44 & 1.38 & 1.30 
& 1.11 & 1.03\\
\hline
\end{tabular}
\end{table*}

The second reduction of the tBESS model we analyze here is defined by $b_L=0$ 
and
$b_R=0.1$. That describes the situation when there are no direct interactions of
the vector resonance to the left top and left bottom quarks. The interaction 
with the right
top quark is at the maximum considered in this paper and the interaction with 
the right
bottom quark can be weakened by the only remaining free parameter $p$.
In Tab.~\ref{Tab:MELDI2}, we show the smallest mass exclusion limits that can 
be found
from the current $WW/WZ$ experimental upper bounds for various values of $g''$ 
when 
$0\leq p\leq 1$. In the table, we also show the value of $p$ and the value of 
the resonance fatness
$\Gamma_\mathrm{tot}/M_\rho$ that correspond to the found mass exclusion limit.
Note that the used experimental upper bounds provide no mass exclusion limits
for $g''\geq 21$.

\begin{table*}[htb]
\centering
\caption{The smallest tBESS resonance mass exclusion limits (MEL)
within the interval $0\leq p\leq 1$ assuming $b_L=0$ and $b_R=0.1$
for various values of $g''$. 
The second and third rows contain, respectively, the values of $p$ and of the 
resonance fatness
that correspond to the quoted mass exclusion limit. 
The excluded masses lie below MEL.}
\label{Tab:MELDI2}
\begin{tabular}{c|cccccccccccccc}
\hline
$g''$ & 14 & 15 & 16 & 17 & 18 & 19 & 20 & \multicolumn{5}{c}{21 -- 25} \\
\hline
$p$ & 1 & 0.871 & 0.772 & 0.707 & 0.672 & 0.630 & 0.589 & 
\multicolumn{5}{c}{--} \\
$\Gamma_\mathrm{tot}/M_\rho$ 
& 0.38 & 0.20 & 0.14 & 0.09 & 0.06 & 0.04 & 0.03 & \multicolumn{5}{c}{--} \\
MEL (TeV) & 2.46 & 2.14 & 2.02 & 1.87 & 1.68 & 1.49 & 1.33 & 
\multicolumn{5}{c}{no MEL}\\
\hline
\end{tabular}
\end{table*}

\begin{figure}[htb]
\centerline{%
\includegraphics[width=7.5cm]{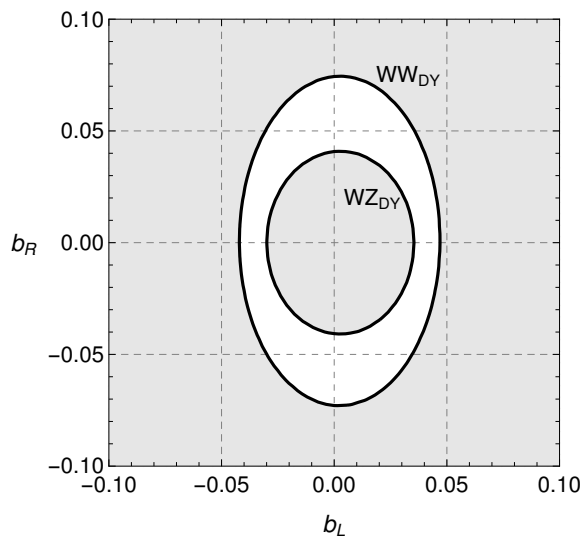}}
\caption{
The excluded region of the $(b_L,b_R)$-plane (gray area) for the vector 
resonance with the mass of 
$M=1.8\unit{TeV}$, $g''=18$, $p=0.8$.}
\label{Fig:2DbLbRlimits}
\end{figure}

To achieve a better understanding of this complex multi-parameter situation we 
can plot the 
regions of the $(b_L,b_R)$ parameter subspace for which a certain resonance 
mass is excluded 
by the experimental upper bounds when the values of $g''$ and $p$ are also 
fixed. As an 
example, we choose $M_\rho=1.8\unit{TeV}$, $g''=18$ and $p=0.8$. With this 
choice
of the parameters the resonance fatness amounts to $\Gamma_\mathrm{tot}/M_\rho 
= 6\%$
at $b_L=b_R=0$. As both, $|b_L|$ and $|b_R|$ approach $0.1$ the fatness grows 
to $9\%$.
Thus we can expect that the deviations introduced by the NWA calculations are 
reasonably small.
The resulting plot of the experimentally excluded regions in the $(b_L,b_R)$ 
space
is shown in Fig.~\ref{Fig:2DbLbRlimits}. 
The excluded region is obtained by the union of the areas excluded by the 
$WZ_\mathrm{DY}$ 
and $WW_\mathrm{DY}$ decay channels. 

The ring-like structure depicted in Fig.~\ref{Fig:2DbLbRlimits} can be 
understood
from the no-direct-interaction graphs of Fig.~\ref{Fig:MELNDI} and from their 
response
to the direct interactions being turned on. With no direct interactions, the 
only channel
that excludes the $1.8\unit{TeV}$ vector triplet is the $WZ_\mathrm{DY}$ one.
However, the tBESS cross section of this channel decreases with the direct 
interactions 
strength. When the direct interactions become sufficiently strong the predicted 
value
dives below the experimental one and the channel ceases to exclude the 
resonance.
On the other hand, owing to the partonic bottom quark contribution, the 
$WW_\mathrm{DY}$ 
decay channel cross section grows with the direct interaction strength. When 
the direct
interactions are turned off the resonance is not excluded by the channel. 
However,
its predicted cross section exceeds the measured value when the direct 
interactions become 
sufficiently strong and, thus, excludes the given resonance. Since the 
exclusion boundary
of the $WZ_\mathrm{DY}$ channel are closer to the origin $b_L=b_R=0$ than the 
exclusion
boundary of the $WW_\mathrm{DY}$ channel a ring-like region
of the $(b_L,b_R)$ parameter subspace, where the resonance is not excluded, has 
emerged.
The resonance fatness at the $WW_\mathrm{DY}$ boundary is $7\%$.

\begin{figure}[htb]
\centerline{%
\includegraphics[width=8.5cm]{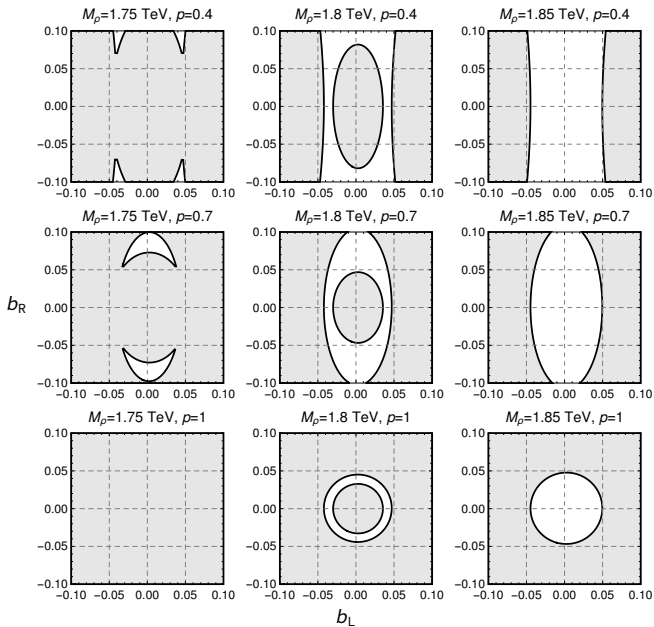}}
\caption{The grid of graphs of the excluded regions of the $(b_L,b_R)$-plane 
(gray area) 
when $g''=18$. The mass changes from graph to graph in the left-to-right 
direction 
through the values of $1.75$, $1.80$, and $1.85\unit{TeV}$. The $p$ parameter 
changes
from the top to the bottom through the values of $0.4$, $0.7$, and $1.0$.}
\label{Fig:2DbLbRlimitsPM}
\end{figure}

The excluded region shown in Fig.~\ref{Fig:2DbLbRlimits} is very sensitive to
the values of the resonance mass. This is demonstrated in 
Fig.~\ref{Fig:2DbLbRlimitsPM}.
The figure consists of the graphs which show the excluded regions
of the $(b_L,b_R)$ subspace for the values of $M_\rho$ that slightly vary 
around the mass
considered at the graph of Fig.~\ref{Fig:2DbLbRlimits}.
The excluded regions change significantly even within the small range 
$1.75\leq M_\rho/\unit{TeV}\leq 1.85$ for all three chosen values of $p=0.4, 
0.7, 1$.
We suggest that the combination of this feature and the sufficiently large
imprecision caused by any used approximation can result in a quite deceiving 
conclusions 
about the exclusion of the resonance of
a given mass. Thus, the deviations
introduced by the NWA calculations should be carefully scrutinized as the 
fatness
of the resonance under consideration grows.

These findings suggest that an exhaustive study of the experimentally 
excluded areas of the complete tBESS parameter space would be a cumbersome task 
even though the 
tBESS phenomenological Lagrangian corresponds to a relatively simplistic LHC 
scenario.
The real scenario of the strongly-interacting extension of the SM can introduce 
even more complex phenomenology of the new resonances to be discovered at the 
LHC.
Even if this is the case the tBESS model analysis can provide valuable 
lessons\footnote{Recently published work \cite{Xie2019} about the LHC signals 
of a model with broad composite resonances can also serve for this purpose.}.
While any simplifying assumptions that would help with the detection analysis 
are certainly appreciated, 
one has to remember that this can come at the price of losing sensitivity
to more subtle behavior of the studied resonances.

\subsection{Flavor physics restrictions of the direct interactions}

Since the direct interaction of the tBESS vector triplet with the third 
generation quarks stands out, it is a source of large flavor violation. In the 
LHC analysis of this paper, the flavor conserving processes dominate. 
Nevertheless, the tBESS resonance direct interaction can significantly 
contribute to the $B-\bar{B}$ mixing, the $B$-meson decays and other flavor 
violating phenomena. Thus, one would wonder how the flavor physics indirect 
limits on the $b_{L,R}$ and $p$ parameters compare with the limits derived from 
the LHC cross section upper bounds discussed in the previous subsection. While 
the exhaustive answer to this question is beyond the scope of this paper we 
would like to provide, at least, a partial insight using the updated versions 
of our previous analyses of the tBESS model.

In the papers~\cite{tBESSprd11,tBESSepjc13}, we studied how flavor physics, 
along with other low-energy measurements, restricts the tBESS model's 
parameters. Among other things, the limits based on the experimental values of 
the $Z\rightarrow b\bar{b}$ decay width and the $B\rightarrow X_s\gamma$ 
branching ratio were derived. Rather than restricting $b_{L,R}$ and $p$ alone, 
the former provides limits on the $b_L-2\lambda_L$ and $p^2(b_R+2\lambda_R)$ 
combinations of the tBESS free parameters and the latter restricts 
$b_L-2\lambda_L$ and $p\cdot(b_R+2\lambda_R)$. Note that the free parameters 
$\lambda_{L,R}$ appearing in these expressions associate with the new 
symmetry-allowed tBESS model interaction terms of the $SU(2)_L$ and $U(1)_Y$ 
gauge bosons. In contrast to the $b_{L,R}$ terms, these terms do not contribute 
to the direct interactions of the vector triplet in the flavor basis. 
Nevertheless, after the mass matrix diagonalization, the physical vertices of 
the electroweak bosons and the tBESS vector triplet with fermions become 
dependent on various combinations of the $b_{L,R}$, $p$, and $\lambda_{L,R}$ 
parameters~\cite{tBESSprd11,tBESSepjc13}. The $\lambda_{L,R}$ parameters did 
not appear in the text above because their impact on the analysis of the direct 
LHC exclusion limits is negligible. However, this is not the case for the 
indirect limits derived from the low-energy measurements. It was also 
found~\cite{tBESSprd11,tBESSepjc13} that these limits vary insignificantly when 
$g''$ changes within the $12\leq g''\leq 25$ region considered in this paper.

\begin{figure}[htb]
\centerline{%
\includegraphics[width=6.5cm]{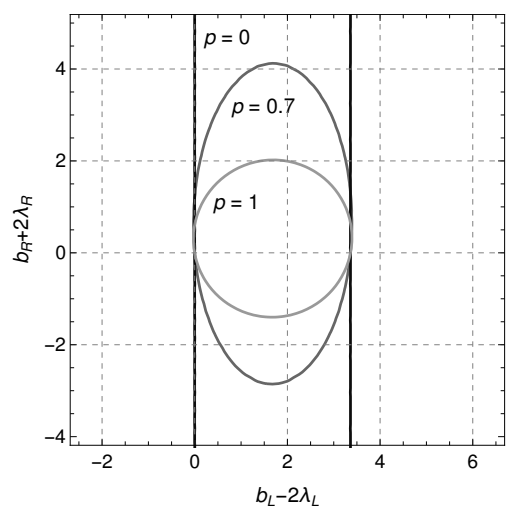}}
\caption{The $95\%$ C.L. $Z\rightarrow b\bar{b}$ allowed regions of the $b$ and 
$\lambda$ parameters
when $g''=18$. The allowed regions form closed elliptical bands;
the darkest gray correspond to $p=0$, the lightest to $p=1$, with
$p=0.7$ in between. When $p=0$, only parts of two parallel bands
of an ``infinite'' ellipse can be seen.}
\label{Fig:DWZbb}
\end{figure}

The examples of the regions of the parametric space allowed by the experimental 
values of $\Gamma_Z = (2.4952\pm 0.0023)\;\mbox{GeV}$
and $\mbox{BR}(Z\rightarrow b\bar{b}) = (15.12\pm 
0.05)\%$~\cite{ParticleDataGroup} for three different values 
of $p$ are displayed in Fig~\ref{Fig:DWZbb}. There, we can see that the allowed 
values of $b_L-2\lambda_L$ and $b_R+2\lambda_R$ are highly correlated and form 
a narrow strip of a circular ($p = 1$) or elliptical ($p = 0.7$) shape. As $p$ 
approaches
zero the shape of the strip straightens up leaving no restriction on 
$b_R+2\lambda_R$. The widths of the strips are about $0.015$ at the $95\%$ C.L. 
The regions displayed in Fig~\ref{Fig:DWZbb} were calculated for 
$g''=18$. However, as mentioned before, the dependence of the obtained limits 
on $g''$ is weak. 
Roughly speaking, $b_L-2\lambda_L$ can assume any value between $-0.01$ and 
$3.5$ independently of $p$. The allowed values of $b_R+2\lambda_R$ range 
between $-1$ and $2$ when $p = 1$. This interval grows as $p$ decreases.

\begin{figure}[htb]
\centerline{%
\includegraphics[width=8cm]{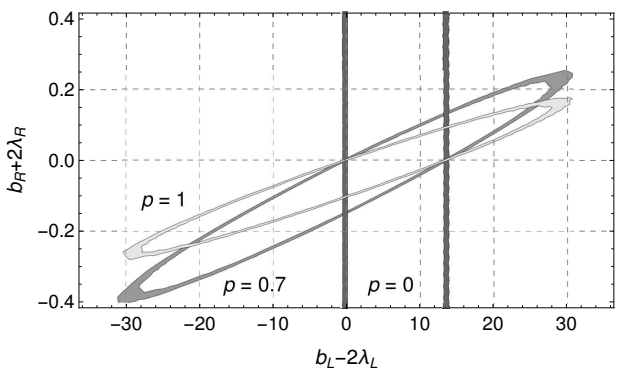}}
\caption{The $95\%$ C.L. $B\rightarrow
X_s\gamma$ allowed regions of the $b$ and $\lambda$ parameters
when $g''=18$. The allowed regions form closed elliptical bands;
the darkest gray correspond to $p=0$, the lightest to $p=1$, with
$p=0.7$ in between. When $p=0$, only parts of two parallel bands
of an ``infinite'' ellipse can be seen.}
\label{Fig:BRb2sgamma}
\end{figure}

The experimental value of $\mbox{BR}(B\rightarrow X_s\gamma) = (3.32\pm 
0.15)\times 10^{-4}$~\cite{ParticleDataGroup,HeavyFlavorGroup} results in the 
allowed regions depicted in Fig~\ref{Fig:BRb2sgamma}. Fig~\ref{Fig:BRb2sgamma} 
shows just three examples of the allowed 
regions in the $(b_L-2\lambda_L, b_R+2\lambda_R)$ parameter space. These 
correspond to $p = 0$, $0.7$, and $1.0$ while $g''=18$. Again, the dependence 
on $g''$ can be ignored for our purpose. The allowed values are highly 
correlated. The $95\%$ C.L. widths of the strips are about $0.005$, $0.007$, and 
$0.7$ for $p=1$, $0.7$, and $0$, respectively. The allowed values of 
$b_L-2\lambda_L$ range between about $-30$ and $30$ when $p = 1$. The compact 
formulation of the allowed values for the remaining parameters read 
$|p(b_R+2\lambda_R)|\lesssim 0.3$.

The combination of the two sets of the allowed regions significantly reduces 
the limits. For example, when $p = 1$ there are four distinct overlaps of the allowed 
regions resulting in the following restrictions. 
First, there is the allowed region that includes the no-direct interactions point:
$|b_L-2\lambda_L|\lesssim 0.01$ and $|b_R+2\lambda_R|\lesssim 0.004$. The second
region is in the vicinity of $b_L-2\lambda_L\approx 0.02$ and 
$b_R+2\lambda_R\approx -0.10$. Further, when $b_L-2\lambda_L\approx 
3.34$ then $b_R+2\lambda_R\approx -0.08$. Finally, there is an allowed region 
around $b_L-2\lambda_L\approx 3.36$ and $b_R+2\lambda_R\approx 0.025$.

There are no low-energy limits on the values of the $b$ and $\lambda$ 
parameters individually. Thus, in principle, $b$'s and $\lambda$'s can be tuned 
to any values if their sum/difference falls within the allowed interval. 
However, if one does not wish to admit a fine-tuning of the parameters, it is 
in place to add some ad hoc restriction; say, the absolute values of the 
$b_{L,R}$ or $\lambda_{L,R}$ parameters should not be greater than 10 times the 
size of the allowed interval for $b_{L,R}\mp 2\lambda_{L,R}$. Of course, if we 
formulated the tBESS model with the simplifying assumption $\lambda_{L,R}=0$ right from the 
beginning then the quoted restrictions would apply directly to the direct 
interaction parameters $b_{L,R}$.

After discussing limits on the $b_{L,R}$ parameters obtained from the LHC upper 
bounds on $\sigma\times \mbox{BR}$ at the end of the previous subsection, the 
text of the present subsection further demonstrates the complexity of the tBESS 
phenomenology and the role as well as the limitations of the flavor physics 
experiments in disentangling it. First of all, the presence of the $\lambda$ 
parameters weakens the restrictions that these measurements can put on the 
direct interactions. Secondly, a thorough study that would include other flavor 
physics processes could help eliminate some of the four regions allowed by the 
combination of the $\Gamma(Z\rightarrow b\bar{b})$ and $\mbox{BR}(B\rightarrow 
X_s\gamma)$ measurements. Having said that, one can see that the flavor physics 
restrictions possess the potential to provide valuable input complementing the 
direct LHC limits. Considering the limited discussion of this subsection only, 
the restrictions shown here are more or less consistent with the limits 
obtained 
from the LHC upper bounds.

\subsection{An experimental update}

In this subsection, we provide a brief review on how the recently published 
experimental 
upper bounds on $\sigma_\mathrm{prod}\times\mbox{BR}$ changed the mass 
exclusion limits
calculated above. These latest upper bounds originate from the ongoing analyses 
of 
the data collected at the LHC experiments. While there have been about 
$139\unit{fb}^{-1}$ of
data recorded to the date the progress of their analysis lags behind. There are
still new bounds being published that are even based on the 2016 dataset of 
$36\unit{fb}^{-1}$.

As expected, even with the bigger integrated luminosity the $WW$ and $WZ$ are 
the only channels 
that provide restrictions on the tBESS production cross sections. In particular,
the new upper bounds for the $WZ$ channel were published in~\cite{R9} and 
\cite{R3}.
The upper bounds were based on the $36\unit{fb}^{-1}$ and $139\unit{fb}^{-1}$ 
datasets, 
respectively. The new upper bounds for the $WW$ channel were published 
in~\cite{R8} and
\cite{R3}. The upper limits of the former paper were based on the 
$77\unit{fb}^{-1}$ dataset.
The new bounds in~\cite{R3} apply to the mass range above $1.3\unit{TeV}$ only.
In combination with some of the previous upper bounds~\cite{43}, these new 
experimental bounds 
result in new mass exclusion limits for the tBESS vector triplet.

To demonstrate the effect of the new upper bounds we update the mass exclusion 
limits
for the scenarios presented in Tables~\ref{Tab:MELNDI}, \ref{Tab:MELDI1}, and 
\ref{Tab:MELDI2}.
Thus, Table~\ref{Tab:MELNDIupdate} contains the updated mass exclusion limits 
when there
are no direct interactions. 
The updated values for the scenarios of Tables~\ref{Tab:MELDI1} and 
\ref{Tab:MELDI2}
are shown in Tables~\ref{Tab:MELDI1update} and \ref{Tab:MELDI2update}, 
respectively.
We can see that the mass exclusion limits have increased by 
more than $1\unit{TeV}$. In addition, all corresponding fatnesses have 
surpassed 
the $10\%$ mark which makes the presented conclusions less reliable. That is why
we restrain ourselves from displaying the mass exclusion limits when they exceed
$3\unit{TeV}$ where the precision of the NWA calculations becomes very 
questionable.

\begin{table*}[htb]
\centering
\caption{The updated mass exclusion limits (MEL) for the tBESS vector resonance 
triplet 
without the direct interactions for various values of $g''$.}
\label{Tab:MELNDIupdate}
\begin{tabular}{c|cccccccccc}
\hline
$g''$ & \multicolumn{5}{c}{12 -- 20} & 21 & 22 & 23 & 24 & 25\\
\hline
$\Gamma_\mathrm{tot}/M_\rho$ 
& \multicolumn{5}{c}{$> 0.40$} & 0.33 & 0.26 & 0.21 & 0.17 & 0.13 \\
MEL (TeV) & \multicolumn{5}{c}{$> 3$} & 2.94 & 2.84 & 2.73 & 2.65 & 2.53$^a$\\
\hline
\end{tabular}\\
$^a$ the mass is also not excluded within $(1.24;1.30)\unit{TeV}$
\end{table*}

\begin{table*}[htb]
\centering
\caption{The updated smallest tBESS resonance mass exclusion limits (MEL)
within the interval $|b_{L=R}|\leq 1$ assuming $b_{L=R}\equiv b_L=b_R$ and $p=1$
for various values of $g''$.
The second and third rows contain, respectively, the values of $b_{L=R}$ and 
of the resonance fatness $\Gamma_\mathrm{tot}/M_\rho$
that correspond to the quoted mass exclusion limit.}
\label{Tab:MELDI1update}
\begin{tabular}{c|cccccccccc}
\hline
$g''$ & \multicolumn{4}{c}{12 -- 19} & 20 & 21 & 22 & 23 & 24 & 25\\
\hline
$b_{L=R}\times 10^2$ & \multicolumn{4}{c}{} & 8.6 & 7.6 & 6.7 & 5.9 & 4.8 & 4.0 
\\
$\Gamma_\mathrm{tot}/M_\rho$ 
& \multicolumn{4}{c}{$> 0.44$} & 0.41 & 0.32 & 0.25 & 0.20 & 0.15 & 0.12 \\
MEL (TeV) & \multicolumn{4}{c}{$> 3$} & 2.96 & 2.86 & 2.74 & 2.65 & 2.52$^a$ & 
2.41$^b$\\
\hline
\end{tabular}\\
$^a$ the mass is also not excluded within $(1.11;1.30)\unit{TeV}$\\
$^b$ the mass is also not excluded within $(1.03;1.30)\unit{TeV}$
\end{table*}

\begin{table*}[htb]
\centering
\caption{The updated smallest tBESS resonance mass exclusion limits (MEL)
within the interval $0\leq p\leq 1$ assuming $b_L=0$ and $b_R=0.1$
for various values of $g''$. 
The second and third rows contain, respectively, the values of $p$ and of the 
resonance fatness
that correspond to the quoted mass exclusion limit.}
\label{Tab:MELDI2update}
\begin{tabular}{c|cccccccccc}
\hline
$g''$ & \multicolumn{4}{c}{12 -- 19} & 20 & 21 & 22 & 23 & 24 & 25 \\
\hline
$p$ & \multicolumn{4}{c}{} & 1 & 1 & 0.962 & 0.910 & 0.796 & 0.755 \\
$\Gamma_\mathrm{tot}/M_\rho$ 
& \multicolumn{4}{c}{$> 0.44$} & 0.40 & 0.32 & 0.25 & 0.20 & 0.14 & 0.11 \\
MEL (TeV) & \multicolumn{4}{c}{$> 3$} & 2.97 & 2.87 & 2.73 & 2.63 & 2.45 & 
2.28\\
\hline
\end{tabular}
\end{table*}

\section{Conclusions}
\label{sec:concl}
Motivated by the absence of any signal of new particles beyond the SM in the 
LHC measurements, we have studied the mass exclusion limits for the 
hypothetical tBESS vector resonance triplet.
The exclusion limits have been established utilizing the experimental upper 
bounds on the $s$-channel resonance production cross section times branching 
ratio provided by the ATLAS and CMS Collaborations for various decay channels. 

The tBESS resonance triplet 
represents a possible signature of a strongly-interacting extension of the SM. 
It has been introduced in the context of the phenomenological 
Lagrangian where, besides the composite $125\unit{GeV}$ Higgs boson, the 
$SU(2)_{L+R}$ triplet of composite vector resonances is explicitly present. The 
vector resonance has been built into the Lagrangian employing the hidden local 
symmetry approach. In the tBESS model, the vector triplet interacts universally 
with all fermions due to the mixing between the electroweak gauge boson fields 
and the vector triplet. In addition, the direct couplings of the vector triplet 
to the top and bottom quarks have been introduced.

Fourteen vector resonance decay channels, for which the experimental upper 
bounds were available to the date, have been considered in our analysis. Of 
these, only the $WW$ and $WZ$ channels provide the mass exclusion limits for the 
tBESS vector triplet in both direct interaction scenarios.

The impact of the direct interactions on the mass exclusion limits has been 
contrasted with the no direct interaction case. As expected, the introduction of 
the specific direct interaction pattern to the model has made the analysis of 
the limits significantly more complex. Besides the emergence of new free 
parameters, the direct interactions have made the bottom quark partonic content 
of the proton significant for the neutral resonance production in the Drell-Yan 
mode. In fact, there 
are the parameter space regions where over $90\%$ of the neutral resonance 
production is comprised of $b\bar{b}\rightarrow \rho^0$. Consequently, the 
sensitivity of $b\bar{b}\rightarrow \rho^0$ to the direct interaction couplings 
impacts all mass exclusion limits founded on the neutral resonance channels 
including the $WW$ one.
The disregard of the bottom quark contents of the colliding protons
would alter our results qualitatively.
The contribution of the partonic charm and strange 
quarks to the neutral and charged vector resonance production is about $5-7\%$.

The experimental upper bounds used in our analysis were derived with the narrow 
resonance qualification. Consistently, the model cross section predictions have 
been calculated in the narrow width approximation. As a rule of thumb we
consider our analysis as reliable for the resonance fatness below $10\%$.
The results obtained for the resonances with the fatness above this mark
must be considered with caution. This is an important issue in the case
of the tBESS vector resonance triplet whose decay width grows significantly with
its mass.

When there are no direct interactions the resonance mass exclusion limits 
range between $2.94\unit{TeV}$ and $2.53\unit{TeV}$ for $g''$ between 21 and 25, 
respectively. The respective resonance fatnesses range between $33\%$ and $13\%$.
Unfortunately, these are already above the $10\%$ rule of thumb. Thus, the quoted limits
should be considered with caution.
When $g''\leq 20$ the mass exclusions limit exceeds $3\unit{TeV}$ and the corresponding
fatness surpasses $40\%$. Therefore, we do not even attempt to quote the particular
exclusion limits obtained by the NWA calculations for $g''\leq 20$.

For the scenario with the direct interactions, the mass exclusion 
limits for two different sets of the parameter constraints have been studied. 
First, it has been assumed that the direct interactions are L-R and top-bottom 
universal, i.e., $b_L=b_R$, $p=1$. In the second case, it has been assumed that 
there are no direct interactions to the left top-bottom quark doublet ($b_L=0$) 
and that the resonance couples strongly to the right top-bottom quark doublet 
($b_R=0.1$). The relative strength of the direct interactions to the top and 
bottom quark has been left as a free parameter. In both cases, the minimal mass 
exclusion limits, when varying the remaining free parameters, have been found. 
In the first case, the limit ranges from $2.96\unit{TeV}$ to 
$2.41\unit{TeV}$ for $g''=20$ and $g''=25$, respectively. 
The respective resonance fatnesses range between $41\%$ and $12\%$.
In the second 
case, the mass exclusion limits for $g''$ between $20$ and $25$ range from 
$2.97\unit{TeV}$ to $2.28\unit{TeV}$, respectively. 
The respective resonance fatnesses range between $40\%$ and $11\%$.
In both cases, 
the mass exclusion limits exceed $3\unit{TeV}$ when $g''\leq 19$.
The corresponding resonance fatnesses surpass $44\%$.

The upper experimental bounds were also used to establish the restrictions 
on the $b_{L,R}$ parameters of the direct couplings of the vector resonance at a given mass. 
It was observed that it was impossible to formulate a universal conclusion about these 
restrictions since the obtained limits strongly depend on the choice of other model's 
parameters, namely the resonance mass and the top-bottom splitting factor $p$. 
As an illustration, when $g''=18$ and $M_\rho = 1.75\unit{TeV}$ the model was completely 
excluded for $p=1$. When  $M_\rho = 1.85\unit{TeV}$ and $p=1$ then $|b_{L,R}| < 0.05$ is allowed. 
When $p$ was lowered to 0.7 the $b_R$ restrictions got relaxed to about $|b_R| < 0.12$ and 
no restrictions on $b_R$ occur when $p=0.4$. Assuming a simplified version of the tBESS model 
where $\lambda_{L,R}=0$ the combination of the $\mbox{BR}(B\rightarrow X_s\gamma)$ and 
$\Gamma(Z\rightarrow b\bar{b})$ measurements provided complementary restrictions on 
$b_{L,R}$ alone. 
With certain qualifications regarding other model's parameter values, it can be concluded 
that these two kinds of restrictions on $b_{L,R}$ neither contradict nor overwhelm each other.
There are non-trivial overlaps of the allowed regions in the parameter space. 
A more comprehensive analysis of how various flavor physics measurements restrict 
the direct interactions would be a worthwhile contribution to the effort of understanding 
of the hypothetical new vector resonances.

\begin{acknowledgements}
We would like to thank P.~Bene\v{s}, F.~Blaschke, and F.~Riva for useful discussions.
The work was supported by the grant LTT17018 of the Ministry of Education, 
Youth and Sports of the Czech Republic and by the COST Action CA15108
"Connecting insights in fundamental physics".
M.G.\ was supported by the Slovak CERN Fund. We would also like to thank the
Slovak Institute for Basic Research for their support.
\end{acknowledgements}

\appendix
 
\section{The tBESS Lagrangian}
\label{app:Lagrangian}

In this Appendix, we review the basic structure of the tBESS effective Lagrangian.
Its phenomenology, relevant to this paper, was discussed in Section~\ref{sec:effLagrangian}.
The complete description and analysis of the Lagrangian can be found 
in~\cite{tBESSprd11,tBESSepjc13,tBESSepjc16}.

The Lagrangian can be split in three parts
\begin{equation}\label{eq:LagTBESS}
  \Lagr = \Lagr_\mathrm{GB} + \Lagr_\mathrm{ESB} + \Lagr_\mathrm{ferm},
\end{equation}
where $\cL_\mathrm{GB}$ describes the gauge boson sector. It includes
the $SU(2)_\mathrm{HLS}$ triplet $\mbox{\boldmath$V$}_\mu$ representing the new
vector resonance triplet.
Further, $\cL_\mathrm{ESB}$ is the scalar sector
responsible for spontaneous breaking of the electroweak and
hidden local symmetries, and $\cL_\mathrm{ferm}$ is the fermion Lagrangian
of the model.

The first term of Eq.~(\ref{eq:LagTBESS}) reads
\begin{eqnarray}\label{eq:LagGB}
   \Lagr_\mathrm{GB} &=& \frac{1}{2g^2}\mathrm{Tr}(\BW_{\mu\nu}\BW^{\mu\nu}) 
   +\frac{1}{2g^{\prime 2}}\mathrm{Tr}(\BB_{\mu\nu}\BB^{\mu\nu}) 
   \nonumber\\ &&
   +\frac{2}{g^{\prime\prime 2}}\mathrm{Tr}(\BV_{\mu\nu}\BV^{\mu\nu}),
\end{eqnarray}
where the field strength tensors of the $SU(2)_L\times U(1)_Y\times SU(2)_\mathrm{HLS}$ gauge fields 
are defined as
\begin{eqnarray}
  \BW_{\mu\nu} &=& \pard_\mu \BW_\nu - \pard_\nu \BW_\mu + \Comm{\BW_\mu}{\BW_\nu},
  \label{Wmunu}\\
  \BB_{\mu\nu} &=& \pard_\mu \BB_\nu - \pard_\nu \BB_\mu,
  \label{Bmunu}\\
  \BV_{\mu\nu} &=& \pard_\mu \BV_\nu - \pard_\nu \BV_\mu + \Comm{\BV_\mu}{\BV_\nu},
  \label{Vmunu}
\end{eqnarray}
where $\BW_\mu=i g W_\mu^a\tau^a$,
$\BB_\mu=i g' B_\mu\tau^3$, and $\BV_\mu=i\frac{g''}{2}V_\mu^a\tau^a$
with the gauge couplings $g,g',\mathrm{\ and\ }g''$, respectively.

The ESB sector contains six unphysical real scalar fields, would-be Goldstone
bosons of the model's spontaneous symmetry breaking.
The six real scalar fields
$\vphi_L^a(x), \vphi_R^a(x),\; a=1,2,3$, are introduced as
parameters of the~$SU(2)_L\times SU(2)_R$ group elements in
the exp-form
$\xi(\vec{\vphi}_{L,R})=\exp(i\vec{\vphi}_{L,R}\vec{\tau}/v)\in SU(2)_{L,R}$
where $\vec{\vphi}=(\vphi^1,\vphi^2,\vphi^3)$.
We express $\Lagr_\mathrm{ESB}$ as a sum of two terms, 
\begin{equation}
\Lagr_\mathrm{ESB}= \Lagr_h + \Lagr_\mathrm{hV}, 
\end{equation}
where $\Lagr_h$ contains the kinetic and mass terms, and the self-interactions of the Higgs boson.
The interaction term
\begin{eqnarray}\label{eq:Lag2}
   \Lagr_\mathrm{hV} &=& -v^2\left[\Tr(\omegaPP)^2(1+2a_V\frac{h}{v})\right.
   \nonumber\\ && \phantom{-v^2[}
    \left.    +\alpha\Tr(\omegaPL)^2(1+2a_\rho\frac{h}{v})\right]
\end{eqnarray}
is responsible for the masses of all gauge bosons including the new vector triplet, and
describes their interactions with the Higgs boson, which are parameterized by
the free parameters $a_V$ and $a_\rho$.

The quantities $\bar{\omega}_\mu^{\parallel}$ and $\bar{\omega}_\mu^{\perp}$
are, respectively, the $SU(2)_{L+R}$ parallel and perpendicular projections of the 
gauged Maurer-Cartan 1-form,
\begin{eqnarray}
   \bar{\omega}_\mu^{\parallel} &=& \omega_\mu^{\parallel}+
   \frac{1}{2}\left(\xi_L^\dagger\BW_\mu\xi_L+\xi_R^\dagger\BB_\mu\xi_R\right)-
   \BV_\mu,
   \label{eq:gaugeMCparallel}\\
   \bar{\omega}_\mu^{\perp} &=& \omega_\mu^{\perp}+
   \frac{1}{2}\left(\xi_L^\dagger\BW_\mu\xi_L-\xi_R^\dagger\BB_\mu\xi_R\right),
   \label{eq:gaugeMCperp}
\end{eqnarray}
where $\omega_\mu^{\parallel,\perp}=
(\xi_L^\dagger\pard_\mu\xi_L\pm\xi_R^\dagger\pard_\mu\xi_R)/2$.

The fermion sector of the Lagrangian can be structured into three parts
\begin{equation}\label{eq:tBESSLagrFerm}
  \cL_\mathrm{ferm} = \cL_\mathrm{ferm}^\mathrm{SM}
                      + \cL_\mathrm{ferm}^\mathrm{scalar}
                      + \cL_{(t,b)}^\mathrm{tBESS},
\end{equation}
where $\cL_\mathrm{ferm}^\mathrm{SM}$ contains
the SM interactions of fermions with the electroweak gauge bosons,
$\cL_\mathrm{ferm}^\mathrm{scalar}$ is about the interactions
of the fermions with the scalar fields and includes the fermion masses.
Finally, $\cL_{(t,b)}^\mathrm{tBESS}$ describes the third generation quark
direct interactions with the vector resonance. In addition, it
contains symmetry-allowed non-SM interactions of
the third quark generation with the EW gauge bosons
parametrized by free parameters $\lambda_L$ and $\lambda_R$.
Here, we show the new physics part of the $(t,b)$ Lagrangian
in the unitary (physical) gauge,
\begin{eqnarray}
  \cL_{(t,b)}^\mathrm{tBESS} &=&
                            i b_L \bar{\psi}_L(\BVslash-\BWslash) \psi_L
  \nonumber\\
                 && + i b_R \bar{\psi}_R P(\BVslash-\BBslash^{R3}) P\psi_R
  \nonumber\\
  & &  + i \lambda_L \bar{\psi}_L(\BWslash-\BBslash^{R3}) \psi_L
  \nonumber\\
  & &  + i \lambda_R \bar{\psi}_RP(\BWslash-\BBslash^{R3}) P\psi_R,
  \label{eq:LagrFermTBESSinUgauge}
\end{eqnarray}
where $\psi$ denotes the usual $SU(2)$ top-bottom doublet and the matrix $P=\diag(1,p)$
disentangles the direct interaction of the vector triplet with the
right top quark from the interaction with the right bottom quark.





\end{document}